\documentclass{mn2e}

\usepackage{epsfig}
\usepackage{latexsym,subfigure}

\def \LCDM {$\Lambda$CDM}

\def \mstar {M$^\star$\,} 
\def \lstar {L$^\star$\,} 

\def \mbjlogh {$\rm M_{b_{\rm J}} - 5 log_{10} h$}
\def \bjm {\rm b_{\rm J}}

\def \kms{{km~s$^{-1}$}}
\def \invkms{{s~km$^{-1}$}}
\def \hkpc{\mbox{$h^{-1}\;\rm kpc$}}
\def \mpc {$h^{-1} {\rm{Mpc}}$}
\def \msol {$\rm M_{\odot}$}
\def \hmsol {$\rm h^{-1} M_{\odot}$}

\def \etal  {\rm {et al.} \rm}
\def \munichsam  {Springel \etal~(2001)} 
\def \munichsambrak  {Springel \etal~2001} 
\def \vdbpapa  {van den Bosch \etal~(2004)}
\def \vdbpapb  {van den Bosch \etal~(2005)}

\newcommand{\plotthree}[6]            
    {\centering \epsfig{file=#1,width=#2\textwidth,clip=}
         \hfill \epsfig{file=#3,width=#4\textwidth,clip=}
         \hfill \epsfig{file=#5,width=#6\textwidth,clip=}}
\newcommand\gsim{\mathrel{\hbox{\rlap{\hbox{\lower4pt\hbox{$\sim$}}}\hbox{$>$}}}}

\begin{document}
\title[Massive Dark Matter Halos around Bright Isolated Galaxies in the
2dFGRS]
{
\vspace{-0.75cm}Massive Dark Matter Halos around Bright Isolated
Galaxies in the 2dFGRS
} 
\vspace{-0.5cm}

\author[Norberg, Frenk \& Cole]{
\parbox[t]{\textwidth}{
\vspace{-1.0cm}
Peder Norberg$^{1,2,3}$, Carlos S. Frenk$^{1}$ \& Shaun Cole$^{1}$ 
}
\vspace*{6pt} \\
$^1$Institute for Computational Cosmology, Department of Physics, University of Durham, South Road, Durham DH1 3LE, UK.\\
$^2$ETHZ Institut f\"ur Astronomie, HPF G3.1, ETH H\"onggerberg, CH-8093 Z\"urich, Switzerland.\\ 
$^3$SUPA\thanks{The Scottish Universities Physics Alliance}, Institute for Astronomy, University of Edinburgh, Royal Observatory, Blackford Hill, Edinburgh, EH9 3HJ, UK.
\vspace*{-0.5cm}}

\maketitle 
 
\begin{abstract}
We identify a large sample of isolated bright galaxies and their
fainter satellites in the 2dF Galaxy Redshift Survey (2dFGRS). We
analyse the dynamics of ensembles of these galaxies selected according
to luminosity and morphological type by stacking the positions of their
satellites and estimating the velocity dispersion of the combined set.
We test our methodology using realistic mock catalogues constructed
from cosmological simulations. The method returns an unbiased estimate
of the velocity dispersion provided that the isolation criterion is
strict enough to avoid contamination and that the scatter in halo mass
at fixed primary luminosity is small. Using a maximum likelihood
estimator that accounts for interlopers, we determine the satellite
velocity dispersion within a projected radius of 175~\hkpc. The
dispersion increases with the luminosity of the primary and is larger
for elliptical galaxies than for spiral galaxies of similar b$_{\rm
J}$ luminosity. Calibrating the mass-velocity dispersion relation using
our mock catalogues, we find a dynamical mass within 175~\hkpc\ of $\rm
M_{175}/h^{-1}M_\odot \simeq 4.0^{+2.3}_{-1.5}
\times 10^{12}\,(L_{\bjm}/L_*)$ for elliptical galaxies and $\rm
M_{175}/h^{-1}M_\odot \simeq 6.3^{+6.3}_{-3.1} \times
10^{11}\,(L_{\bjm}/L_*)^{1.6}$ for spiral galaxies. Finally, we
compare our results with recent studies and investigate their
limitations using our mock catalogues. 
 
\end{abstract}


\section{Introduction} 

The view that galaxies are surrounded by large dark matter halos dates
back more than 30 years to the pioneering study of the rotation curve
of M32 by Rubin \& Ford (1970). Extended galactic halos are, in fact,
a generic feature of the cold dark matter model of galaxy formation
(Blumenthal et al 1984, Frenk et al 1985), but this fundamental
theoretical prediction has limited observational support.  Zaritsky
\etal~(1993) attempted to measure the mass and extent of dark matter
halos by analysing the dynamics of satellite galaxies found around
`isolated' galaxies. Since galaxies generally have only a few
detectable satellites, they used a method that consists of
stacking satellites in a sample of primaries of similar luminosity.
In spite of the small size of their relatively inhomogeneous sample,
Zaritsky \etal~(1993) were able to detect massive halos around
isolated spiral galaxies extending to many optical radii. Having
nearly doubled their satellite sample to 115 members, Zaritsky \etal
(1997b) confirmed their earlier claims including a puzzling lack of
correlation between the velocity dispersion of the (stacked) satellite
system and the luminosity of the primary.

More recently, McKay \etal~(2002) performed a similar analysis on data
from the Sloan Digital Sky Survey (SDSS; York \etal~2000). They
compared mass estimates derived from satellite dynamics to those
derived from weak lensing analyses of the same data (McKay \etal
2001). With a much larger sample than that of Zaritsky et al (1997),
they were able to detect a correlation between satellite velocity
dispersion and primary luminosity. This trend was confirmed by Prada
\etal~(2003) who also used SDSS data. Although they are both based on
SDSS data, these two studies find 
results that, while consistent at first sight, are in, fact, somewhat
contradictory. 
For example, 
although Prada et al (2003) found a strong dependence of satellite
velocity dispersion on galactrocentric distance, their measured
velocity dispersion within a radius of 125~\hkpc\ is similar to the
values obtained by McKay \etal~(2002) at a radius of
275~\hkpc. Discrepant results were also found by Brainerd \& Specian
(2003) who applied the same technique to the early, ``100k'' data
release of the 2degree-Field Galaxy Redshift Survey (2dFGRS; Colless
2001) and derived satellite velocity dispersions which are in
qualitative and quantitative disagreement with those of Zaritsky \etal
(1997), McKay \etal~(2002) and Prada \etal~(2003). A more extensive
analysis of the complete 2dFGRS (Colless \etal~2003) by Brainerd
(2005) also led to disagreements with the results of earlier work.
This somewhat confused picture of satellite dynamics is due in large
part to different choices of primary galaxy samples and to differences
in the modelling and analysis methods.

This paper has multiple aims. Firstly, we carry out a new analysis of
the dynamics of satellites around bright galaxies of different
morphological types selected from the full 2dFGRS. The goal is to
constrain the velocity dispersion and mass of their dark matter halos
and we therefore select a sample of isolated galaxies chosen according
to strict criteria. Secondly, we investigate the reliability and
accuracy of commonly used dynamical analysis methods. For this, we
make extensive use of realistic mock catalogues constructed from large
cosmological N-body simulations and different semi-analytic galaxy
formation models (Cole \etal~2000; \munichsambrak). A similar
approach, but in a different context, was adopted by
\vdbpapa. Finally, we attempt to understand the root cause of the
differences found in previous work, again relying on the use of
realistic mock catalogues.

The paper is organised as follows. In Section~\ref{sec:data}, we
briefly present some of the characteristics of the 2dFGRS data and
simulations used for our analysis. In Section~\ref{sec:sample}, we
describe the satellite sample selection scheme, together with its
robustness to changes in the selection parameters. The analysis
of the 'stacked' satellite velocity distribution is carried in
Section~\ref{sec:anal} while, in Section~\ref{sec:vel_disp_est}, we
present velocity dispersion estimates for our mock catalogues and for
2dFGRS primaries split according to luminosity and morphological
type. Using a model for the relationship between dark halo mass and
satellite velocity dispersion, we give, in Section~\ref{sec:mass_est},
an estimate of the mass of the halos around 2dFGRS galaxies. In
Section~\ref{sec:discussion}, we compare our results with those of
previous studies, and we conclude in Section~\ref{sec:conclusion}.

\section{The data}
\label{sec:data}

\subsection{The 2\lowercase{d}FGRS data}
\label{sec:data.2df}

Detailed descriptions of the construction of the 2dF Galaxy Redshift
Survey (2dFGRS) and its properties are given in Colless \etal~(2001;
2003).  In summary, galaxies are selected down to a magnitude limit of
$b_{\rm J}\approx\,19.45$ from the full 2dFGRS catalogue, with
$\sim 225\,000$ galaxies having a ``good quality'' redshift measurement. We
restrict our analysis to the two large contiguous volumes of the
survey, one centred on the Southern Galactic Pole (SGP) and the other
close to the direction of the Northern Galactic Pole (NGP).

Three limitations of the 2dFGRS catalogue are relevant for this work.
Firstly, the 2dFGRS source catalogue, based on the APM galaxy
catalogue, is not complete.  By comparing the 2dFGRS with the SDSS,
Norberg \etal~(2002) estimated the completeness of the 2dFGRS to be
$\sim 91\% \pm 2\%$, and ascribed the incompleteness primarily to
misclassification of APM images. Misclassification of close galaxy
pairs will cause some true pairs to be missed from our sample of
primaries. The 2dFGRS suffers from an additional form of close pair
incompleteness due to `fibre collisions' during the spectroscopic
observations. The observing strategy employed in the 2dFGRS,
consisting of a set of overlapping tiles which are successively
observed, was designed to minimise the number of `fibre collisions',
and the remaining incompleteness is very precisely quantified (Colless
\etal~2001).  Finally, the {\it rms} accuracy of redshift measurements
for a typical galaxy is 85~\kms\ and tends to be slightly larger for
the faintest galaxies and slightly smaller for the brighter ones (Colless
\etal~2001). 

\begin{figure} \centering
\epsfig{file=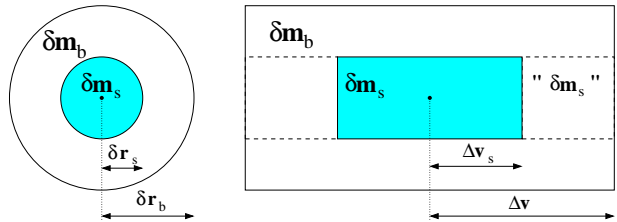,width=0.47\textwidth,clip=,} \caption{A
schematic of the isolation criterion. The figure on the left shows the
projection on the sky of the cylinder that defines the primary
isolation criterion. The figure on the right shows the configuration as
seen along the redshift axis. The central point in each panel
corresponds to a primary; only galaxies located in the shaded area are
considered as potential satellites.}  \label{fig:cyl} \end{figure}

Since we are primarily interested in the velocity dispersion at large
radius, the close pair incompleteness in the catalogue  does not have an
important effect on our conclusions. Nevertheless, it is important to
model carefully both the incompleteness and the velocity errors and to
include these in the construction of our mock galaxy catalogues.

Finally, we make use of the `eyeball' morphological classification
carried out by Loveday \etal~(1996) based on the APM images. This is
available for 80\% of the central galaxies in our sample. 
We prefer
this classification to the more objective spectral classification of
Madgwick \etal~(2002) because aperture effects are important for our
sample of relatively nearby galaxies. However, we have repeated the
analysis of section~\ref{sec:vel_disp_2df} using subsamples defined by
their spectral classification and find no difference in our results
within the errors.

\subsection{The \LCDM\ Simulation}

We use mock 2dFGRS catalogues constructed from cosmological
simulations in order to assess the extent to which the various
limitations of the data affect our results. In particular, we use the
mocks to investigate possible systematic effects arising from our
method for selecting satellites, as well as from our procedure of
stacking satellites together. The mock catalogues allow us also to
investigate the effects of redshift space distortions and redshift
measurement errors. In the simulations we, of course, know not only
the redshifts of galaxies but also their distances. In what follows,
we use the term ``real space'' to refer to measurements that make use
of the true 3-d position and the term ``redshift space'' to refer to
measurements that make use of pseudo 3-d positions, i.e. those for which
the distance to the galaxy is given by the sum of the pure Hubble flow
distance and the peculiar velocity along the line-of-sight, in units of
\mpc.

\begin{table*}
\caption{Properties of the combined NGP \& SGP satellite samples 
around bright galaxies, for different values of $N_{\rm viol}$ and for 
different primary morphological type. The numbers quoted in brackets  
are after small groups are excluded from the full satellite sample. 
$r_{\rm p}$ is in~\hkpc.}
\begin{tabular}{ccccccccc}   
\hline
Sample & Primary Type & $N_{\rm \rm viol}$ &
\multicolumn{2}{c}{$N_{\rm prim}$} &  \multicolumn{2}{c}{$N_{\rm
sat}$} & \multicolumn{2}{c}{$N_{\rm sat}(r_{\rm int} < r < r_{\rm
ext}) $} \\ 
& & &  &  &  &  & $\phantom{4}0<r_{\rm p}<175$ & $40<r_{\rm p}<175$ \\
\hline
2dFGRS & any & 0  & 362 & (357) & 642 & (588) & 273 & 241   \\
2dFGRS & any & 4  & 571 & (564) & 1003 & (918) & 434 & 383   \\
2dFGRS & Spiral-Irregular & 0 & 203 & (203) & 322 & (322) & 141 & 120  \\
2dFGRS & Spiral-Irregular & 4 & 311 & (311) & 465 & (465) & 210 & 181   \\
2dFGRS & Elliptical-S0 & 0 & \phantom{1}85 & (\phantom{1}81) & 203 &
(161) & \phantom{1}75 & \phantom{1}69  \\ 
2dFGRS & Elliptical-S0 & 4 & 141 & (135) & 338 & (265) & 118 & 111   \\
Mock & any & 0 & 387 & (387) &  648  & (648) & 384 & 231   \\ 
Mock & any & 4 & 736 & (736) & 1226 & (1226) & 723 & 442   \\ 
\end{tabular}
\label{tab:stats}
\end{table*}

To construct mock 2dFGRS catalogues, we use a high-resolution N-body
simulation of a flat, $\Lambda$-dominated cold dark matter universe
with the following parameters: matter density, $\Omega_{\rm m}=0.3$;
cosmological constant term, $\Omega_{\Lambda}=0.7$; Hubble constant,
$H_{0}=70~$km s$^{-1}$Mpc$^{-1}$; index of primordial fluctuation
power spectrum, $n=1$; and present-day fluctuation amplitude
$\sigma_8=0.9$. The simulation followed 400$^3$ particles in a box of
of side 110~\mpc\ (see Gao \etal~(2004) for a full description of the
simulation).  The simulation was populated with galaxies by applying
the ``Munich'' semi-analytic model of galaxy formation to the merger
trees of each halo (\munichsambrak). In this model, galaxies reside in
resolved halos and their subhalos. When a subhalo is no longer
resolved, the galaxy is placed on the most bound particle of the
subhalo when it was last resolved and an analytic dynamical friction
calculation is used to determine when a satellite merges with the
central galaxy. 
The free parameters of the model are tuned to match
the Tully-Fisher relation, the B-band cluster galaxy luminosity
function and the overall two-point correlation function (Springel
\etal~2001). 
An interesting feature of this semi-analytic galaxy
formation model is the generation of a velocity bias between galaxies
and dark matter, as function of halo radius. 

In order to obtain as close a match as possible between the 2dFGRS
selection function and that in the mocks, we rescale the luminosities
of the model galaxies 
preserving the luminosity ranking 
so that their
luminosity function exactly matches that of the 2dFGRS (Norberg
\etal~2002). The required rescaling can be as large as a magnitude for
some of the mock galaxies, but the differential rescaling for galaxies
brighter than
\mbjlogh$<-18$, which make up the bulk of our sample of primaries, is
small.  Thus, for these brighter galaxies, the magnitude differences
are quite accurately preserved. We then extract magnitude-limited
catalogues of galaxies to the same magnitude limit and with the same
geometry as the real 2dFGRS, as described by Norberg \etal~(2002).
This requires using three periodic replications of the simulation
cube. (We have checked that removing any duplicated systems
does not affect any or our results.)

The last step is to create `sampled' mocks by applying the 2dFGRS
masks (including the redshift incompleteness mask) and 
the 2dFGRS photometric errors,
as explained in greater detail in Norberg \etal~(2002). Finally, we add
to each observed velocity an 
``observational error'' randomly sampled from a Gaussian of width
80~\kms. This value is a compromise between the Gaussian errors
measured from repeat observations for our sample of 2dFGRS primaries
($\sigma \simeq 70 \pm 7$~\kms) and satellites ($\sigma \simeq 86 \pm
5$~\kms).

\section{Satellite Sample}
\label{sec:sample}

We begin this section by explaining the method used to define the
satellite sample, which is applied to both the 2dFGRS data and the
mock catalogues. We then briefly consider the robustness of the
satellite properties to variations of the selection parameters, an
issue that we address further in the appendices.  Finally, we present
some general properties of the satellite samples used in this paper.

\subsection{Satellites around Isolated Primaries}
\label{sec:algo.data}

Since the main purpose of our analysis is to constrain the mass of the
galactic halo using the dynamics of satellites, we require a sample of
isolated primaries. To construct it, we begin by excluding regions in
the 2dFGRS that could be contaminated by large clusters. Specifically,
we exclude regions lying within three projected Abell radii ($3\times
1.5$\hkpc) and 3000~\kms\ of the centre of clusters in the catalogue
of Dalton \etal~(1997). Next, we select a sample of bright, isolated
primaries by requiring that they satisfy the following criteria:

\begin{itemize}
\item the local 2dFGRS magnitude limit should be at least $\delta m_s =
  2.2$ fainter than the primary 
\item all the neighbouring galaxies within $\Delta V\,=\,|V^{\rm
    prim}-V^{\rm gal}|\,\le\,2400$~\kms\ and within a projected radius
  $\delta r_s \le 400$~\hkpc\ should be faint enough to satisfy $b_{\rm J}^{\rm gal} -
  b_{\rm J}^{\rm prim} \ge \delta m_s  = 2.2$  
\item all neighbouring galaxies within $\Delta V\,=\,|V^{\rm 
    prim}-V^{\rm gal}|\,\le\,2400$~\kms\ and within a projected radius
  $\delta r_b \le 1000$~\hkpc\ should satisfy $b_{\rm J}^{\rm gal} -
  b_{\rm J}^{\rm prim} \ge \delta m_b = 0.8,$
\end{itemize}
where the projected radius between two galaxies, at positions ${\bf r_1}$ 
and ${\bf r_2}$, is defined by:
\begin{eqnarray}
\delta d \,=\,2\,\frac{|{\bf r_1} + {\bf
r_2}|}{2}\,\sqrt{\frac{1-\cos(\alpha)}{1+\cos(\alpha)}} \,\,\,\, {\rm
with} \, \cos(\alpha)\,=\,\frac{{\bf r_1}\cdot{\bf r_2}}{|{\bf
r_1}|\,|{\bf r_2}|} \,\, . 
\label{eq:proj_radius}
\end{eqnarray}

All galaxies that lie within a projected distance $\le 400$~\hkpc\ of
a primary and have relative velocity $\Delta V_s\,=\,|V^{\rm
prim}-V^{\rm gal}| \le 1200$~\kms\ are considered as potential
satellites. The isolation criterion is presented in schematic form in
Fig.~\ref{fig:cyl}. 

Our adopted value of $\delta m_s$ corresponds to a factor of 8 in
luminosity and a similar factor in mass. The primary motivation behind
this choice is to ensure that the satellites are small enough to
produce only minor perturbations in the gravitational potential of the
system. As shown in section~\ref{sec:vel_disp_est}, our galaxy mock
catalogues indicate that our adopted value is adequate. A more
detailed discussion of the precise choice of $\delta m_s$ may be found
in Appendix~A.

Since not all galaxies in the 2dFGRS have a measured redshift, the
primary isolation criterion could be violated by galaxies that lack a
redshift measurement. 
We can guard against this by eliminating all primaries that could
have their isolation criterion violated by such galaxies. 
A less restrictive condition is to accept only those for which it could
be violated by at 
most N$_{\rm viol}$ galaxies. 
A conservative estimate of $N_{viol}$
follows from taking all galaxies without redshift to be at the
redshift of the primary and checking whether this would cause the
primary to violate the isolation criterion. 
In what follows, we adopt
a value of N$_{\rm viol} = 4$.  We have checked that none of our
results are influenced by the precise choice of N$_{\rm viol}$
used. The only effect of adopting N$_{\rm viol} = 4$, rather than
N$_{\rm viol} = 0$,  is to increase
the number of satellites and hence the signal-to-noise of our
measurements. With this choice of N$_{\rm viol}$, the satellite sample
is $\sim$~55~\% larger than for N$_{\rm viol}=0$ ($\sim$~45~\% for
spirals and $\sim$~65~\% larger for ellipticals). The largest increase
occurs for the faintest absolute magnitude bins for which the number
of satellites nearly doubles.

With this algorithm and using the values of the selection parameters
specified above, we identify 571 primary galaxies surrounded by 1003
satellites. The algorithm also detects over 1500 isolated galaxies
without spectroscopically confirmed satellites brighter than the local
magnitude limit, but these have over 2200 neighbours without measured
redshift which could in principle be satellites. Some of these
statistics are summarised in Table~\ref{tab:stats}.

Applying the same selection criterion to the mock catalogues, we
identify, in real space, 750 primary galaxies surrounded by 1241
satellites. In real space, the isolation criterion is slightly
different: instead of two cylinders of length $\Delta V$ and $\Delta
V_s$, we use two spheres of radius $\delta r_b$ and $\delta r_s$
respectively. The requirements on the magnitude differences remain the
same. All galaxies within $\delta r_s$ of the primary are considered
as satellites. In redshift space, including velocity errors, we find
736 primary galaxies surrounded by 1226 satellites. As for the real
data, the algorithm detects a further 1500 primaries without any
confirmed satellite, but with 1700 possible candidates without
measured redshifts.

\subsection{Robustness of the satellite detection algorithm}
\label{sec:algo.rob}

\begin{figure}
\centering
\epsfig{file=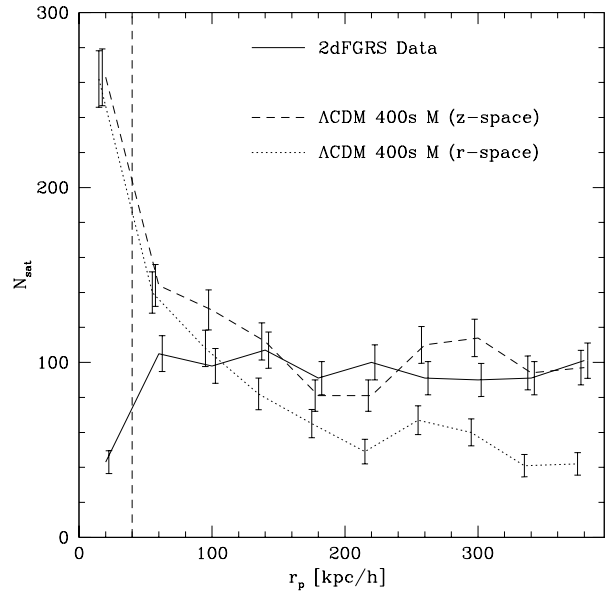,width=0.47\textwidth,clip=,}
\caption{Transverse distributions of satellites: comparison between
  satellite samples extracted from the 2dFGRS (solid bold line), from
  the mocks in redshift space (dashed line) and in real space 
  (dotted line). The vertical dashed line indicates the limiting projected 
  radii, as used for the `r$_{\rm p}$ cut' (see text). The errorbars 
  plotted assume Poisson statistics.}
\label{fig:rad.dist}
\end{figure}

It is important to test the dependency of the satellite detection
algorithm on the values of the selection parameters. We find that the
size of the satellite sample and its properties are not very sensitive
to the specific values of the inner and outer projected radii, so long
as the exclusion criterion does not become too restrictive. The
adopted values for these parameters represent a compromise between
having a dynamically isolated system and a large sample of satellites.

On the other hand, the inner and outer cylinder depths have a
non-trivial influence on the satellite sample. First, if the `velocity
difference' between $\Delta V_s$ and $\Delta V$ is less than $\Delta
V_s$\footnote{i.e. $|\Delta V - \Delta V_s| \le \Delta V_s$}, then
there is the potential risk of finding a single satellite galaxy
associated with two different primaries. In such a  case, neither 
primary can be considered isolated. 
Therefore in order to avoid this problem and obtain a self consistent
isolation criterion, we always impose $|\Delta V - \Delta V_s| \ge
\Delta V_s$. We note that none of the previous work in this
subject has included such a constraint.
It is not clear if their samples had any shared
satellites and if they did how they were treated, but we note that for
both the 2dFGRS data and our mock galaxy catalogues, this subtle
problem does occur.

Secondly, if the outer depth of the cylinder, i.e. $\Delta V$, is 
very large, the isolation criterion is more stringent , and
our catalogue will contain fewer primaries. Conversely if
$\Delta V$  is too small the 
isolation criterion becomes too relaxed and we risk including
non-isolated systems in our sample.

\begin{figure*}
\centering
\epsfig{file=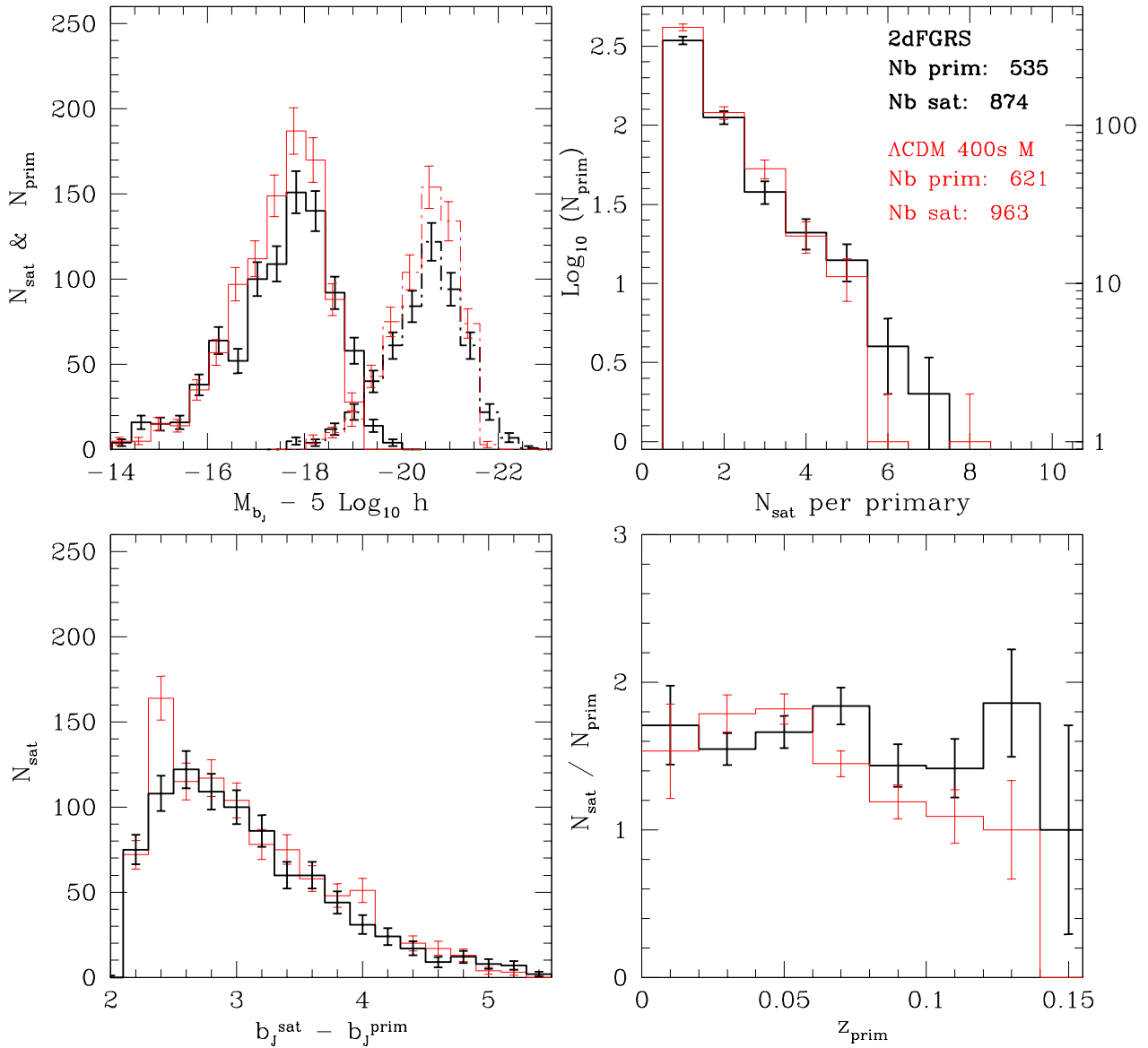,width=0.97\textwidth,clip=,}
\caption{Statistics of primary galaxies and their associated satellites: in all
  panels we adopt a thick line style for 2dFGRS data and a
  thin line style for final mock data. All errorbars assume Poisson 
  statistics. 
  The top left panel shows the distribution of absolute
  magnitudes of satellites (solid line) and primaries (dot-dashed
  line). The top right panel shows the distribution of the number of
  satellites per primary. The bottom left panel shows the distribution
  of the magnitude difference between primary and satellite. The bottom
  right panel shows the fraction of satellites per primary as function
  of redshift. All panels are done using the satellite samples obtained 
  by applying the isolation criterion, rejecting small groups and with 
  `r$_{\rm p}$ cut'. 
  See text for more details.}
\label{fig:gen_prop}
\end{figure*}

Finally, if
$\Delta V_s$ increases the contamination of the satellite sample
by 
interlopers\footnote{In the mock catalogues we can label any galaxy
  that is selected as a satellite, but does not reside within
  the dark matter halo of the primary galaxy as an interloper. In the
  real data this is not possible, but one can still statistically
  estimate the fraction of interlopers by their effect on the
  satellite velocity distribution (see section~\ref{sec:interlopers}).} 
is increased.
For instance, when
increasing $\Delta V_s$ from 600~\kms\ to 1800~\kms,
we see a flattening of the velocity distribution, which we interpret as
being due to interlopers in our satellite catalogue. 
Conversely too small a value of $\Delta V_s$ 
would mean
that the full width of the velocity distribution that we are trying
to characterise would not be sampled. It is essential that
$\Delta V_s$ be greater than 3 to 4 $\sigma$, where $\sigma$ is 
rms width  of the underlying satellite velocity distribution.

Therefore the choice of $\Delta V = 
2400$~\kms\ and $\Delta V_s = 1200$~\kms\ is a compromise between reducing 
the catalogue contamination from interlopers, increasing the size of 
the satellite catalogue and allowing a robust velocity dispersion
estimate to be obtained. We note that the choice of the depth of 
the cylinders could be tuned with the size of the system considered
to increase the efficiency with which `small' systems are detected.
This is a point to which we return in section~\ref{sec:vel_disp_est}.

In the appendices, we discuss in detail the isolation 
criteria proposed by McKay \etal~(2002), Prada \etal~(2003), Brainerd \& 
Specian (2003), \vdbpapa\ and Brainerd (2005). The summary of those 
findings is given in section~\ref{sec:discussion}.

\subsection{General Properties of Satellite Samples} 

Before performing a detailed dynamical study, we focus briefly
on some general properties of the satellite samples. This leads
us to make some additional cuts to improve the match between observed
and mock samples.

\subsubsection{Transverse distributions of satellites}
\label{sec:trans_dist}

In Fig.~\ref{fig:rad.dist}, we present the 
distribution of transverse  separations for
three satellite samples. The mock satellite samples selected both in
real and redshift space have centrally peaked distributions
with the redshift space selected sample having the flatter
distribution at separations $r_{\rm p}>100$~\hkpc.
In contrast the data has a almost flat distribution
throughout $100<r_{\rm p}<350$~\hkpc, but with a significant
drop in the central region.
The reasons behind this difference on small scales are multiple: first of all, 
the 2dFGRS input catalogue lacks close galaxy pairs (Norberg
\etal~2002) and this deficit is enhanced by the targets 
that are rejected due to fibre collisions; secondly galaxies are not point 
source objects, but extended objects on the sky, 
which means that the innermost radial bin can suffer from 
projection effects which are not taken into account in the mocks 
(see \vdbpapb\ for a more detailed study of this particular issue). 
Finally, the large number of satellites in the innermost radial bin 
is not a generic model prediction: their number is sensitive to  
details of the dynamical friction prescription used in the
semi-analytic model.

We could model this limitation in the mock catalogues, by e.g. 
implementing a supplementary incompleteness around each primary. 
However, we preferred not to add an arbitrary incompleteness model to 
our analysis, and hence choose to restrict some comparisons to 
satellite samples with satellites $r_{\rm p}> 40$~\hkpc .
Hereafter we refer to this as the `r$_{\rm p}$ cut'. The vertical 
line in Fig.~\ref{fig:rad.dist} indicates this inner limiting 
projected radius. Beyond this radius, the distributions of
transverse separations for satellites from the mocks
(in redshift space) and from the 2dFGRS
are in approximate agreement.

\subsubsection{Satellite sample properties}

In Fig.~\ref{fig:gen_prop}, we present general properties from the mock 
and data satellite samples (with light and bold line-styles respectively). 
The data and the semi-analytic mock satellite samples have very 
similar properties: the peak of the satellite and primary absolute magnitude 
distributions are roughly the same for both samples, with primaries being 
typically one magnitude brighter than \mstar (top left panel of 
Fig.~\ref{fig:gen_prop}); the fraction of satellites per primary 
does not vary significantly as function of redshift; the overall 
distribution of satellite-primary magnitude differences is similar 
for both samples, as is the shape of the 
distribution of the number of satellites per primary 
(top right panel of Fig.~\ref{fig:gen_prop}).

The good match between data and mock satellite samples, shown in
Fig.~\ref{fig:gen_prop}, is achieved after applying both the `r$_{\rm
p}$ cut' and an additional cut to remove small groups from the 2dFGRS
sample. We have removed all primaries which have 9 or more
satellites. In the mock catalogue there is only one primary with more
than 6 satellites, but in the 2dFGRS there are a few primaries,
satisfying the isolation criterion, with 9 or more satellites. Most of
them, more than 85\%, have primaries that are elliptical galaxies.
Removing these systems reduces our 2dFGRS satellite sample by 85
satellites and reduces the the average number of satellites per
primary from $\sim$~1.75 to $\sim$~1.63.  The reason for removing them
is to both achieve a better match between the mock and 2dFGRS samples
and to avoid our dynamical estimates being dominated by these small
groups. 
We could have achieved this second goal by retaining the
groups, but down weighting them by giving equal weight per primary
rather than per satellite. However, we find that within the errors 
this does not change our results.

In summary, with these extra restrictions, we end up with mock 
satellite samples which are rather similar to the 2dFGRS satellite 
samples. 

\section{Modelling the satellite velocity dispersion}
\label{sec:anal}

\begin{figure}
\centering
\epsfig{file=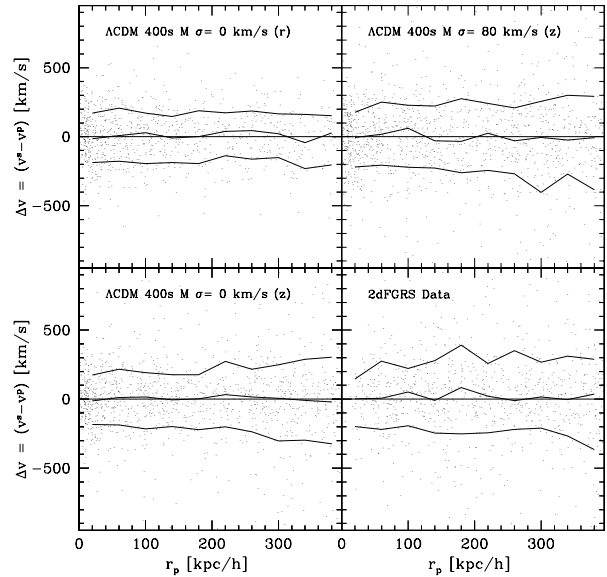,width=0.47\textwidth,clip=,}
\caption{Velocity difference of satellite galaxies and primaries versus
  transverse separation for the 2dFGRS sample (bottom right panel) and for
  the mock samples: real space (top left panel); redshift space without
  velocity errors (bottom left panel) and with velocity errors (top
  right panel). The solid lines show respectively the 16$^{\rm th}$, 
  50$^{\rm th}$ and 84$^{\rm th}$ percentiles of the distribution, in bins of 
  40~\hkpc. This plot has not been corrected for interlopers.}
\label{fig:vel_rp}
\end{figure}

In this section, we consider the dynamical properties of the
satellite samples obtained in section~\ref{sec:sample}. We start by 
looking at the satellite velocity distributions, then address the 
issue of interlopers and background subtraction. Once they are well 
understood, we devise a method to estimate the satellite 
velocity dispersion of stacked primaries.

\begin{figure}
\centering
\epsfig{file=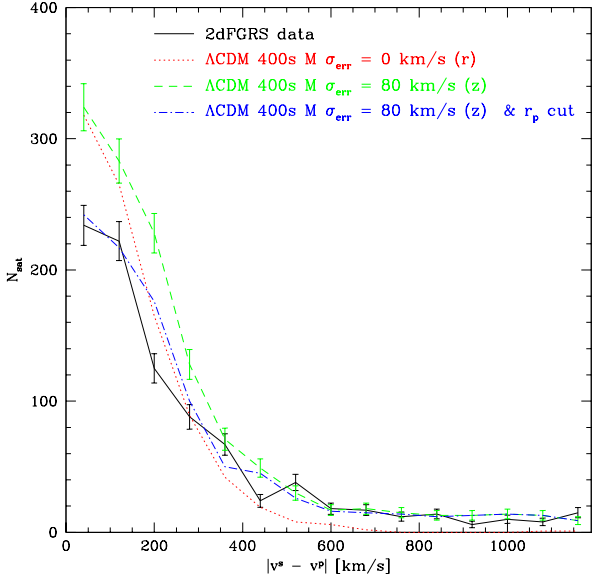,width=0.47\textwidth,clip=,}
\caption{Velocity distributions of satellites within a projected radius
  of 400~\hkpc: comparison between 2dFGRS satellite sample (solid bold
  line) and the mock samples in real and redshift space (with and
  without `r$_{\rm p}$ cut'). Errorbars, only plotted on two of the curves 
  for visibility, assume Poisson statistics. See text for discussion.}
\label{fig:veldist_sat}
\end{figure}

\subsection{Velocity Distribution of Satellites}

Fig.~\ref{fig:vel_rp} is a scatter plot showing, for the 2dFGRS data
and the mock samples, the satellite galaxy velocity difference (with
respect to its associated primary) versus its projected distance 
from the primary. In all panels, the velocity distribution is rather 
symmetric around zero for all projected distances. The precise 
choice of the cylinder depth (as fixed by $\Delta V$ and $\Delta V_s$)
does not have a strong influence on the distribution for either 
data or mock, except that the 16$^{\rm th}$  and 84$^{\rm th}$ 
percentiles become slightly more noisy as the sample size is reduced. 
There is no strong correlation between velocity difference 
of the satellite-primary pair and the satellite's projected distance 
from the primary. We note that Fig.~\ref{fig:vel_rp} has not been 
corrected for interlopers, an issue we address in 
Section~\ref{sec:interlopers}.

\begin{figure}
\centering
\epsfig{file=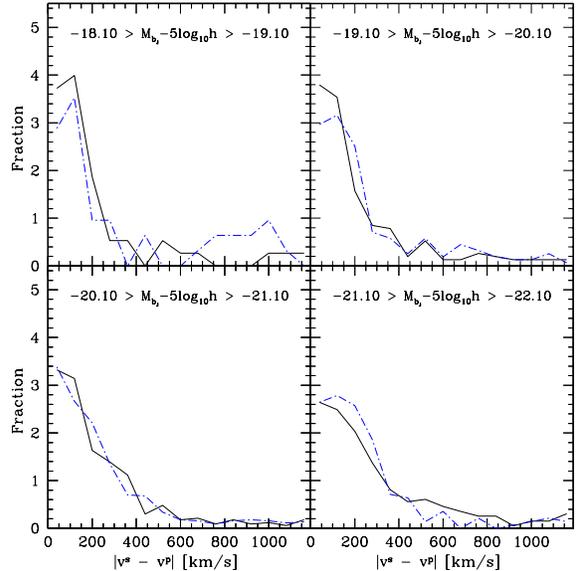,width=0.47\textwidth,clip=,}
\caption{Velocity distributions of satellites within a projected radius
  of 400~\hkpc: similar plot as Fig.~\ref{fig:veldist_sat} (with same
  lines types showing 2dFGRS data and mock satellite sample in redshift 
  space, with velocity errors and `r$_{\rm p}$ cut' applied), but split 
  by absolute magnitude (indicated in each panel). The y-axis is here
  divided by the total number of satellites in each sample, to
  facilitate comparisons between samples of different sizes. 
  For clarity, we omit errorbars. See text for discussion.}
\label{fig:veldist_sat_magbin}
\end{figure}

In Fig.~\ref{fig:veldist_sat}, we consider the velocity distribution 
of satellites averaged over projected radii.
Comparing real and redshift space samples, we see that the velocity
errors and the redshift space distortions tend to broaden the
velocity distribution. Applying the `r$_{\rm p}$ cut' to the
redshift space mock slightly alters the shape of the velocity distribution
(the number of satellites with smaller 
velocity difference is reduced more than those with larger velocity 
difference). 
The dependence of velocity dispersion on project radius seen in the
mock might not occur in the real Universe. If so the incompleteness
at small $r_{\rm p}$ has no effect and the 2dFGRS satellite samples can 
be interpreted as sampling the full velocity distribution within 
a given radius. However if the real satellites are like those in the
mock we should restrict our comparisons to large scales where 
the `r$_{\rm p}$ cut' has no influence.

The velocity distribution of the mocks after applying the 
`r$_{\rm p}$ cut' is close to the one measured from 
the 2dFGRS. Both sets of satellite catalogues have
a velocity distribution with an extended 
tail, with a nearly constant amplitude beyond $\sim$~800~\kms .
This is to be expected
due to contamination from interlopers and needs to be dealt with 
when analysing redshift space distributions.

At this stage we split the samples by
absolute magnitude, as our theoretical prejudices, leads us to
expect that the satellite velocity distributions will be more extended around
brighter primaries. This is exactly what we see in the various panels
of Fig.~\ref{fig:veldist_sat_magbin}.
The two top panels show the velocity distributions around the faintest 
primaries. For both samples, it seems clear that a velocity limit 
of 1200~\kms\ is probably too large. Although not plotted the real
space samples do not contain any galaxy with velocity 
$\Delta V\gsim$~450~\kms). 
Moreover, the faintest mock sample contains a `lump' of galaxies 
at $\Delta V \sim 900$~\kms. This indicates that 
for faint primaries a smaller value for $\Delta V_s$ should be
chosen, as otherwise contamination from interlopers will be very strong.
The two bottom panels show the satellite velocity distributions around the 
brighter primaries. They both show the existence of satellite galaxies
with large relative velocities, especially the brightest sample for
which $\sim$~20\% ($\sim$~5\%) of the satellites have a relative
velocity larger than 500~\kms\ (900~\kms). Therefore in order to measure
the velocity dispersion of these systems it is essential to sample 
the full width of the velocity distribution. 

\begin{figure*}
\centering
\epsfig{file=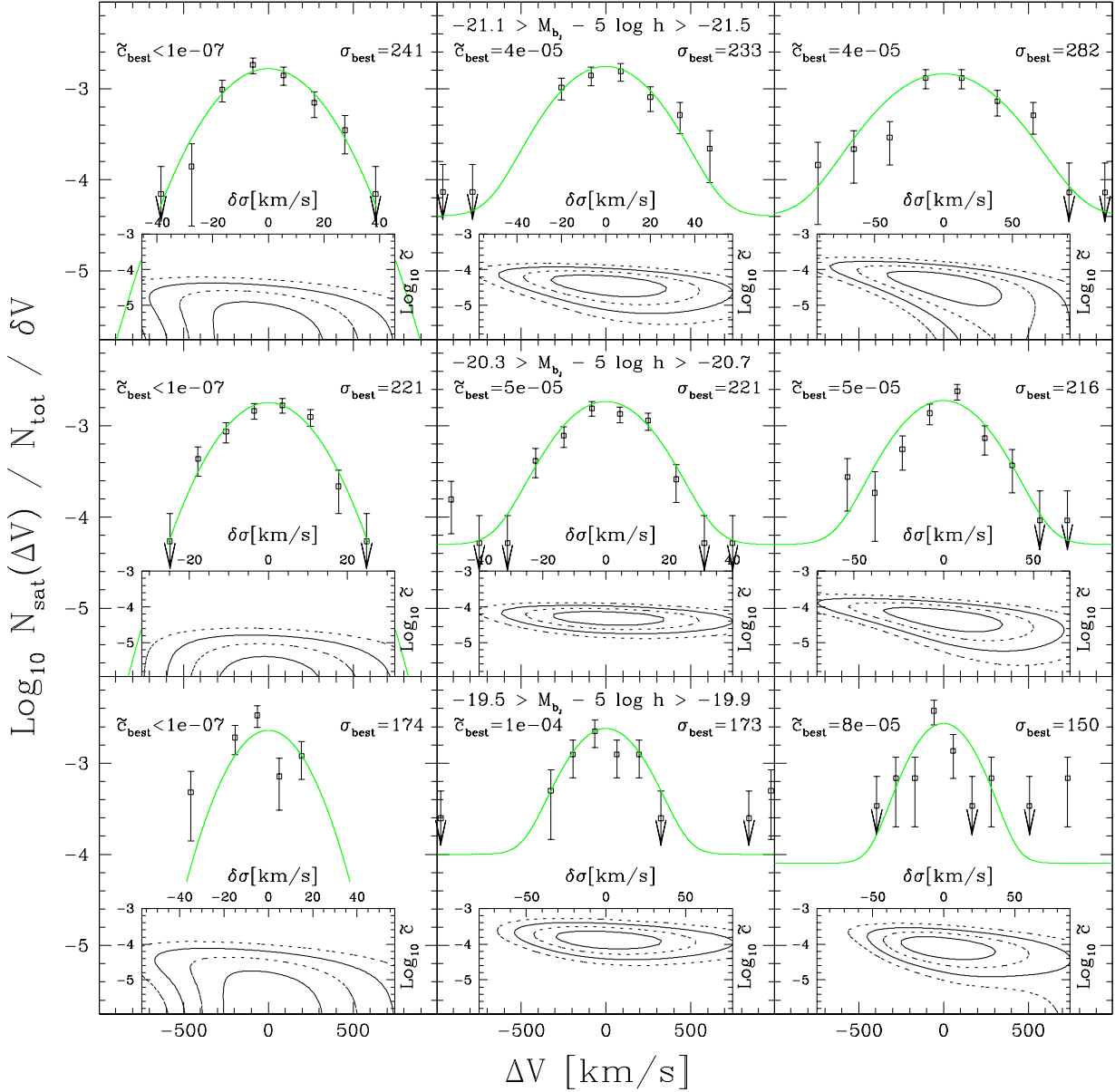,width=1.00\textwidth,clip=,}
\caption{Fitting a Gaussian plus constant to the velocity distribution
  of satellite galaxies around primaries of three different absolute 
  magnitude
  bins (bright to faint from top to bottom, with magnitude range
  indicated in the middle column) taken from respectively the
  mock satellite samples in real and redshift space and from the 2dFGRS
  data (from left to right). The binned satellite velocity distribution 
  is shown by the squares with errorbars (assuming Poisson statistic) 
  and the smooth curve the best fitting Gaussian plus constant, as 
  determined by solving Eq.~\ref{eq:likelihood}. The binning used is 
  regular and of width 3/4~$\sigma_{\rm best}$, which, given in each panel, 
  is the best velocity dispersion estimate, expressed in \kms. Intervals 
  without bins 
  represent velocity intervals without any satellites in. The best fit 
  constant, $\tilde{c}$, is also given in \invkms.  The inset 
  in each panel show, in dotted, the 1 and 2-$\sigma$ contour levels, in 
  the $\delta \sigma - {\rm log_{10}\,\tilde{c}}$ plane, of the two 
  parameter fit and, in solid, the corresponding one parameter 
  confidence contours. We point out that the velocity dispersion error 
  contours are relative to $\sigma_{\rm best}$ and that insets in 
  different panels have different scales. See text for further details.}
\label{fig:fit_3magbin}
\end{figure*}

We note that, with the exception for the brightest sample,
the velocity distributions of the mocks are quite similar 
to those extracted from the 2dFGRS for each absolute 
magnitude split sample. For the brightest sub-sample the
2dFGRS velocity distribution is wider than that of the mock.
We find that this is related to the presence bright
`isolated' ellipticals which are found in 2dFGRS sample.
We address the influence of the morphological mix in 
section~\ref{sec:vel_disp_2df}.

\subsection{Interlopers and Background Subtraction}
\label{sec:interlopers}

In the previous section, we came across one important concern for 
the satellite sample: its contamination by interlopers.  The core 
of the velocity distributions shown in
Figs.~\ref{fig:veldist_sat} and~\ref{fig:veldist_sat_magbin} are
rather well described by Gaussian distributions, with some extended 
tails. These tails are due to interlopers, i.e. galaxies
which in redshift space just happen to come within the selection region,
but which in real space are more distant and belong to another 
halo/system.

Our stacked systems are probing a cylindrical volume in redshift
space. 
Hence to first order we would expect
the interlopers to be randomly distributed within this cylinder.
This motivates modelling the velocity distribution of
each stacked system as the sum of a  Gaussian and a constant.
More precisely we set up a maximum 
likelihood estimator based on the following probability function:
\begin{eqnarray}
{\rm p}(v) & = & \frac{f(v)}{\int_{-v_{\rm fit}}^{+v_{\rm fit}}\,f(v)\,{\rm d}v } \,\,\, ,\\
f(v) & = & \frac{1}{\sqrt{2\pi}\sigma}\,{\rm\,exp}\left(-\frac{v^2}{2\,\sigma^2}\right)\,+\,{\rm  \tilde{c}}\,\,\, ,
\label{eq:prob_dist}
\end{eqnarray}
where the velocity fitting range is between -$v_{\rm fit}$ and 
$v_{\rm fit}$. Hence the maximum likelihood estimator is just the 
product of the probabilities ${\rm p}(v_{i})$ associated with each 
satellite, which can be written as: 
\begin{eqnarray}
{\rm ln}[\mathcal{L}] & = & -2\,\sum_{i=1}^{N}\,{\rm
  ln}\left[f(v_{i})\right]\,\nonumber\\
 & & +\,2\,N\,{\rm ln}\left[1-{\rm erfc}\left(\frac{v_{\rm
  fit}}{\sqrt{2}\,\sigma}\right)\,+\,2\,v_{\rm fit}\,{\rm \tilde{c}}\right] \,\, .
\label{eq:likelihood}
\end{eqnarray}

By maximising this likelihood as function of $\sigma$ and $\rm \tilde{c}$ 
for each sample of stacked primaries, we are able to determine a 
typical velocity dispersion for these systems, together with the 
fraction of interlopers. 
We note that this approach can be applied 
to any subsample of satellite galaxies. Hence, we can, for example, 
test for a radial dependence of the satellite velocity dispersion 
by just considering satellites in different projected radial 
shells.
The $\sigma$ we measure in this way will
be the underlying velocity dispersion of the stacked satellite system
added in quadrature with the rms error of the measured
satellite-primary velocity difference. 
This simple subtraction would only be invalid if the 
measured satellite-primary velocity differences were correlated
with the measurements errors and there is no evidence for this
in the 2dFGRS.

In Fig.~\ref{fig:fit_3magbin}, we show the results of this method for
three magnitude bins (bright to faint from top to bottom) taken from
the mock satellite samples in real (left) and redshift space (middle) and
from the 2dFGRS data (right hand column). Note that we have 
not subtracted in quadrature the
`known' velocity
measurement error, which for the data and the simulation is
$v_{\rm err}\sim\,110$~\kms\ (see  section~\ref{sec:vel_disp_2df}),
from the fitted velocity dispersion.
The inset in each panel shows the 
1 and 2-$\sigma$ confidence regions\footnote{Assuming Gaussian errors, 
the likelihood  given by Eq.~\ref{eq:likelihood} is distributed like 
a $\chi^2$ distribution with two degrees of freedom 
($\rm \sigma, \tilde{c}$).}: as dotted contours for two free parameters 
(i.e. $\Delta\,\chi^2\,=\,2.30,\,6.17$) and as solid contours for one free 
parameter (i.e. $\Delta\,\chi^2\,=\,1.0,\,4.0$). As
expected, the real space samples (i.e. left most
column of Fig.~\ref{fig:fit_3magbin}) are all fit by a pure Gaussian,
as in real space we do not have the problem of interlopers. 
The interloper fraction averaged over the fitted range is given by
\begin{eqnarray}
{\rm I (\sigma, \tilde{c})} & = & \frac{2\,\tilde{c}\,v_{\rm fit}}{\int_{-v_{\rm fit}}^{+v_{\rm fit}}\,f(v)\,{\rm d}v } 
\label{eq:int_frac}
\end{eqnarray}
and for the samples in redshift space we find a roughly constant value.
For galaxies in the range -20.3~$\ge$~\mbjlogh~$\ge$~-20.7, and assuming 
a cylinder depth of 1200~\kms, we find interloper fraction 
of $10.7^{+3.7}_{-2.4}$\%. Interestingly the interloper fraction in the 
mock samples and the 2dFGRS data are similar for all the three 
magnitude bins presented in Fig.~\ref{fig:fit_3magbin}. This is another 
example of how well our mock catalogues mimic the real data.

Finally, we note that the size of the error on the satellite velocity
dispersion differs between the mock and 2dFGRS samples. The typical
1-$\sigma$ uncertainty on $\sigma$ for the 2dFGRS data is
approximately 30~\kms, independent of the best fitted velocity
dispersion, whereas for the mock samples uncertainty is closer to
20~\kms.  This is probably related to the greater homogeneity of mock
catalogues, which, despite their high level of sophistication, do not
contain as much variety as the real 2dFGRS data (see
section~\ref{sec:vel_disp_2df}).  

\begin{figure}
\centering
\epsfig{file=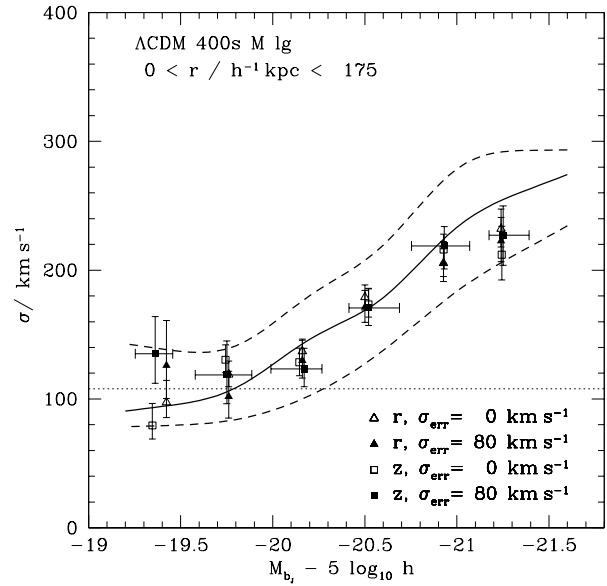,width=0.47\textwidth,clip=,}
\caption{Velocity dispersion of `stacked' primaries of given absolute
  magnitude for various mock samples: in real space with and without
  velocity errors included (filled and open triangles respectively) and 
  in redshift
  space with and without velocity errors included (filled and open 
  squares respectively). The errorbars plotted are the 1-$\sigma$ errors as
  obtained from the 2 parameter fits on samples taken with cylinder
  depths of $\delta V = 1200$~\kms, for samples brighter than \mstar, and 
  with cylinder depths of $\delta V = 600$~\kms, for the fainter samples. 
  The solid line is the median satellite velocity
  dispersion, as estimated from the volume limited semi-analytic galaxy
  catalogue. The two dashed lines shows the 16$^{\rm th}$ and 
  84$^{\rm th}$ percentiles of the velocity dispersion distribution. For 
  samples where the 80~\kms\ velocity uncertainty has been included, 
  we subtract from the estimated velocity dispersion the total velocity 
  uncertainty, $v_{\rm err}$, in quadrature. See text for discussion.}
\label{fig:sig.mag}
\end{figure}

\section{Velocity Dispersion of Satellite Systems}
\label{sec:vel_disp_est}

We can now estimate the velocity dispersion of satellites
around stacked primaries 
for different ranges of absolute magnitude within a chosen 
limiting transverse radial separation. The limiting radius
needs to be large enough so that the composite 
satellite system contains sufficient satellites, but small enough 
so as to only sample dynamically connected regions. From theoretical 
considerations, a limiting radius of 175~\hkpc\ is reasonable for 
halo masses between $5 \times 10^{11}$ and 10$^{13}$~\hmsol, as it is smaller 
than their typical virial radius, but still large enough to sample 
a fair fraction of the virial volume. Unless otherwise specified,
our measurements are all done within a projected separation 
of 175~\hkpc\ from the primary.

\subsection{Mock Satellite Velocity Dispersion}
\label{sec:vel_disp_mock}

In Fig.~\ref{fig:sig.mag}, we plot the estimated velocity dispersion from 
the mock satellite samples in real (triangles) and redshift (squares) 
space, with and without velocity errors included (filled and empty 
symbols respectively). All are in pretty good agreement with the 
distribution of velocity dispersions measured directly from the full 
semi-analytic simulation cube, whose median and 
associated 16 and 84 percentiles are shown by the solid and
dashed curves respectively. 
We note that when the volume limited simulation cube is analysed in
this way 
including or excluding the most central satellites has a systematic effect on 
the measured velocity dispersion. Discarding all satellites within 
a projected radius of $\sim$~5~\hkpc\ of the primary, results in
a velocity dispersion which is systematically larger, by 5 to 15\%.

\begin{figure}
\centering
\epsfig{file=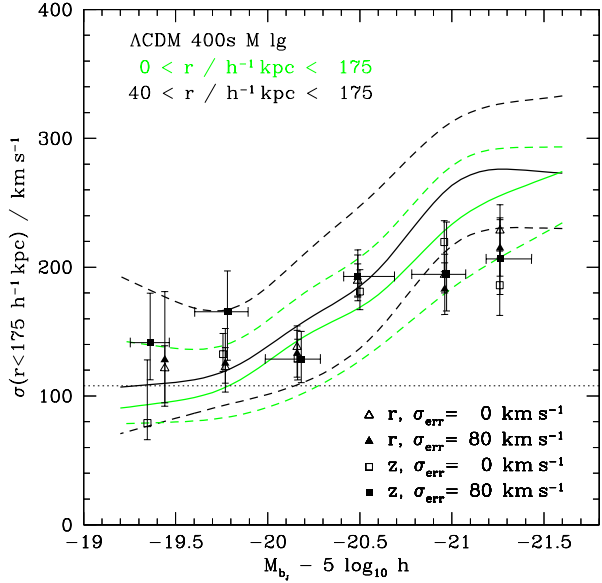,width=0.47\textwidth,clip=,}
\caption{Velocity dispersion of `stacked' primaries of given absolute
  magnitude for various mock samples: same as Fig.~\ref{fig:sig.mag}, 
  but with the `r$_{\rm p}$ cut' applied. In light grey, we plot, as for 
  reference, the same lines as shown in Fig.~\ref{fig:sig.mag}.}
\label{fig:sig.mag_rpcut}
\end{figure}

Fig.~\ref{fig:sig.mag_rpcut} is like Fig.~\ref{fig:sig.mag}, but 
with the `r$_{\rm p}$ cut' applied to the samples analysed from
both the simulation cube and mocks.
For this reason, the black solid and 
dashed curves in Fig.~\ref{fig:sig.mag_rpcut} are slightly different 
to those in Fig.~\ref{fig:sig.mag} (reproduced in grey in 
Fig.~\ref{fig:sig.mag_rpcut}). The agreement between the various 
velocity dispersion estimates is not as good as 
in Fig.~\ref{fig:sig.mag}. Nevertheless,
the satellite velocity dispersion inferred after applying 
the `r$_{\rm p}$ cut' agrees within the 
(typically 50\% larger) errors with those 
measured before this cut was applied. 
In other words, the velocity dispersion as measured from satellite galaxies is 
not too sensitive to the innermost spatial distribution. This is 
probably not a big surprise, because of the mixed selection effects
that come 
into the isolation criterion. Despite the less than perfect agreement between 
the simulation and mock in Fig.~\ref{fig:sig.mag_rpcut}, it is
important to have demonstrated that the
velocity dispersion inferred from satellite galaxies is 
in general agreement with the median halo velocity dispersion 
measured in the simulation. However, the fact
that the agreement is not perfect should also not be forgotten when
interpreting the results of dynamical studies of stacked 
satellite systems.

We recall that to arrive at the velocity dispersion estimates plotted
as the filled triangles in Figs.~\ref{fig:sig.mag_rpcut}
and~\ref{fig:sig.mag} we have subtracted in quadrature 
the mean velocity measurement error $v_{\rm err}\sim$~110~\kms, which
comes from adding in quadrature the velocity measurement errors of
primaries and the satellites.  This has the desired effect of
producing estimates that are all in reasonable agreement with each
other, but also has the consequence that the fractional error on
the estimated velocity dispersion is increased by a factor
$\sigma/(\sigma^2-v_{\rm err}^2)^{1/2}$ relative to that of the
$\sigma$ that comes from the maximum likelihood
algorithm. Nevertheless, Figs.~\ref{fig:sig.mag}
and~\ref{fig:sig.mag_rpcut} show that for nearly the full absolute
magnitude range covered by our data we are able to recover the
underlying velocity dispersion without any strong biases due to
selection effects.

\subsection{Satellite Velocity Dispersion from 2dFGRS}
\label{sec:vel_disp_2df}

\begin{figure}
\centering
\epsfig{file=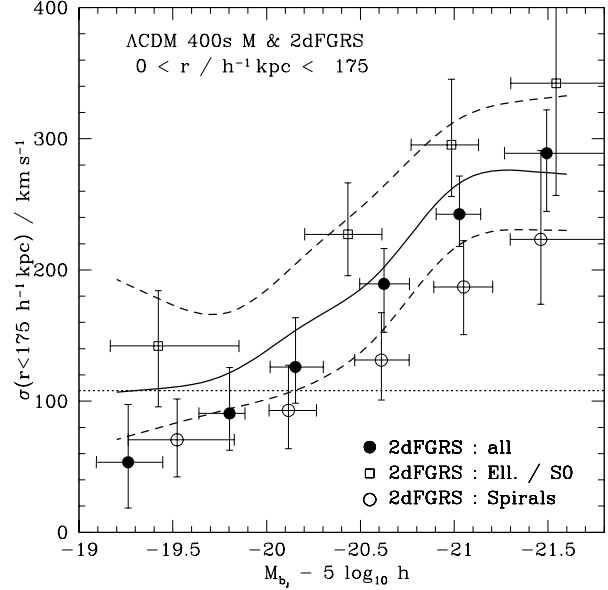,width=0.47\textwidth,clip=,}
\caption{Velocity dispersion as measured using the isolation criterion
  as function of associated primary absolute magnitude. The data sample is
  shown for all primary types (filled solid circles), but also split by 
  primary galaxy morphology: elliptical-S0 (open squares) and 
  spiral-irregulars (open circles). The solid and dashed lines, same lines as 
  shown in Fig.~\ref{fig:sig.mag_rpcut}, are plotted here for 
  comparison purpose only. See text for discussion.}
\label{fig:sigma_amag_morph}
\end{figure}

For the 2dFGRS satellite sample, we find it useful to split the sample
based on whether the primary is an isolated spiral or elliptical galaxy.
In Fig.~\ref{fig:sigma_amag_morph}, we plot the velocity dispersions
for the combined 2dFGRS samples (filled circles) as well for the subsets of
satellites with  spiral (and irregular) primaries 
(open circles) and  elliptical-S0 primaries (open squares).
In the mocks, if we infer the morphological types of the primaries
from the bulge-to-disk ratios, then we also find a small fraction of 
elliptical primaries. However, the correspondence between bulge-to-disk
ratio and morphology is crude and so we prefer not to split the mock
sample in this way. 
Fig.~\ref{fig:sigma_amag_morph} shows that
the velocity dispersions from the mocks are in very good agreement with
those from the overall 2dFGRS sample, but we also see in the 2dFGRS
that the velocity dispersions of satellites around
spiral primaries are significantly lower
than those of satellites around elliptical primaries. 

We also note that for the 2dFGRS the typically error on the estimated
velocity dispersion is twice as large as in the corresponding mock
sample.  This is most certainly related to a combination of the
following factors: the mock satellite samples are typically two times
larger than the two 2dFGRS satellite samples (as the mock catalogues
used are not split by morphology); due to the limited physics
included the mock samples are likely to be more statistically
homogeneous than the real data; the average number of satellites per
primary is slightly smaller in the data than in the mocks and the
smaller total number of satellites in the data will lead to more
scatter in the estimated velocity dispersion.

Based on the correspondence between the estimated satellite velocity
dispersions and those of the underlying dark matter haloes found for
the mocks in Section~\ref{sec:vel_disp_mock}, 
we make the claim that the velocity
dispersions of the primaries as inferred from the 2dFGRS satellite
sample should be a reliable tracer of the primaries `true' velocity
dispersions. 
Quantitatively, we expect this method to work for
`isolated' galaxies brighter than \mbjlogh~$\simeq\,-19.0$, i.e. galaxies 
which are as bright or brighter than the Milky-Way.

\section{Mass estimate of isolated systems}
\label{sec:mass_est}

In section~\ref{sec:vel_disp_est}, we showed that it is possible, over
a range of absolute magnitudes, to recover from the satellite 
velocity dispersion the underlying halo velocity dispersion. 
Therefore, it is tempting to go one step further and
infer the mass which is dynamically enclosed in these systems. There are,
nevertheless, many issues which needs to be dealt with in order to 
obtain a reliable mass estimate. Two major concerns are, of course, which mass 
estimator to use and how to choose the radius within which to measure
the mass. Our approach is to calibrate a mass estimator using the 
mocks, as for them we know the mass of the parent dark matter halo of each
galaxy.

\begin{figure}
\centering
\epsfig{file=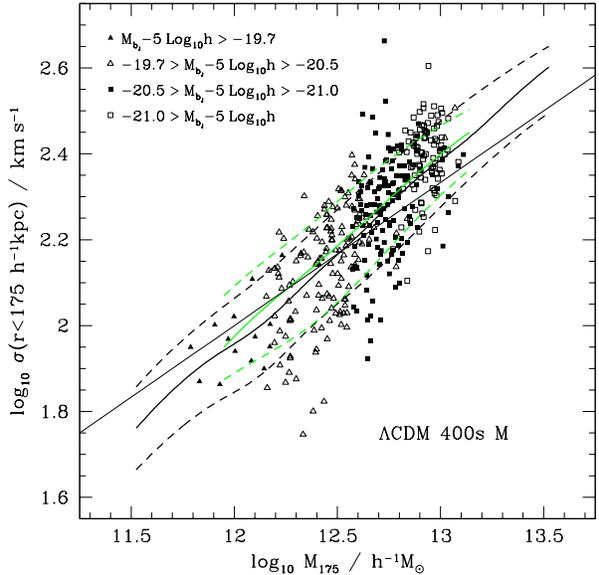,width=0.47\textwidth,clip=,}
\caption{Satellite velocity dispersion in the volume limited mock
  catalogue as function of the associated primary halo mass. Both the halo mass and the 
  satellite velocity dispersion are measured within a radius of 175~\hkpc.
  The different symbols correspond to isolated systems found in the mock 
  survey and labelled as function of primary luminosity (see key). 
  The thick grey and black solid lines correspond to the median satellite 
  velocity dispersion as measured from the simulation cube for all primaries 
  found in the mocks and all central galaxies in the simulation cube 
  respectively. The associated dashed line corresponds to 16$^{\rm th}$ and 
  84$^{\rm th}$ percentiles of the satellite velocity dispersion distributions.
  For comparison purpose the median dark matter velocity dispersion is
  also plotted as a thin black line.
  }
\label{fig:veldisp_mass}
\end{figure}

\subsection{Calibration of mass estimator}

The way we calibrate the halo mass-luminosity relation is to 
measure in the simulation the relation between halo mass and halo
velocity dispersion and so obtain a way to relate the measured velocity 
dispersion to a halo mass. In Fig.~\ref{fig:veldisp_mass}, we plot the 
satellite velocity dispersion as measured in the volume limited 
semi-analytic catalogue as function of the associated halo mass within 
175~\hkpc, M$_{175}$. The varying symbol type indicates the 
absolute magnitudes of the isolated primaries (see the figure legend).
The galaxy formation 
model used preserves very accurately the 
mass-luminosity hierarchy of central galaxies\footnote{This is only true 
for halo masses with circular velocities below 400~\kms, as for 
halos with larger circular velocities, an artificial cut-off in the cooling 
recipe creates much fainter central galaxies in the very largest halos.}. 
By this we mean not only that on average bright central galaxies reside in 
massive halos and faint central galaxies in the less 
massive ones, but also that the scatter in this whole mass-luminosity relation 
is quite small. This is an essential assumption which needs 
to be accurately satisfied both in the model and the genuine data
for the method of stacked primary systems to work. 

Mocks created from the same dark matter simulation, but with the 
Cole \etal~(2000) galaxy formation model have a much larger scatter in 
the relationship between halo mass and central galaxy luminosity and can 
therefore not be used in the calibration process. The scatter present 
in those mocks does not allow satellites 
of primaries of similar luminosity to be stacked, as they can belong to 
systems which are intrinsically too different in mass.
Since this work began, a whole new 
suite of semi-analytic galaxy formation models with AGN feedback have appeared 
(e.g. Croton \etal~(2005), Bower \etal~(2006)). These models
also have quite a large scatter in the relationship between
halo mass and primary luminosity, though less than in the
Cole \etal~(2000) semi-analytic model.
We simply restate that by stacking systems by primary luminosity
and then attempting to infer the halo mass, we are implicitly assuming
that the scatter between luminosity and mass is small and so
in calibrating such a relation we should use a mock catalogue
in which this is true.

Using \munichsam\ semi-analytic model of galaxy formation
in the full simulation cube, we show in 
Fig.~\ref{fig:veldisp_mass} that using all central galaxies or just
those whose primaries satisfy the isolation criterion make very little
difference to the relation satellite velocity dispersion and dark
matter halo mass.
In both cases the medians and the 
16$^{\rm th}$ and 84$^{\rm th}$ percentiles of the satellite velocity 
dispersion distributions match well over one full magnitude in halo 
mass. They are both well parametrised by a power-law relation of the 
form $\sigma_{175} \propto (M_{\rm 175})^\alpha$, 
with $\alpha$ ranging between $\sim$~0.42 and $\sim$~0.56, depending on which 
percentile of the velocity dispersion distribution one attempts to fit. 
We note that in our mock we do not find any isolated systems 
residing in halos outside the range $\sim 5 \times 10^{11}h^{-1}$~M$_{\odot}$
 $\sim 10^{13}h^{-1}$~M$_{\odot} $. 
Hence any mass estimate outside this range is based on assuming
the good correspondence found in Fig.~\ref{fig:veldisp_mass} around 
isolated halos and central galaxies holds over a larger mass range.
Finally, we have investigated whether the calibrating relation in the
mocks is the same for both elliptical and spiral primaries. For our crude
bulge-to-disk ratio assignment of morphological type we find the relations
are virtually identical and so we have opted to use the one overall
relation present in Fig.~\ref{fig:veldisp_mass}
in all cases.

\begin{figure}
\centering
\epsfig{file=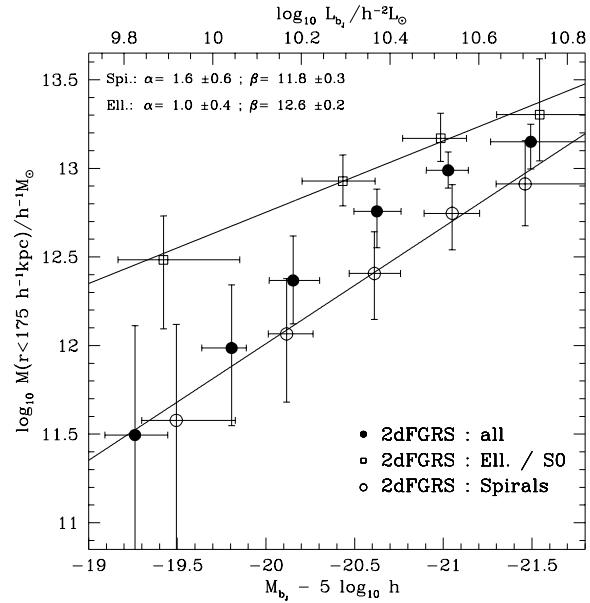,width=0.47\textwidth,clip=,}
\caption{Simulation calibrated halo mass, within 175~\hkpc, as function 
  of primary absolute magnitude. The open squares corresponds to the 
  mass of the halos of elliptical galaxies, whereas open circles 
  to spiral halo masses. The solid dots corresponds to the halo mass of 
  an `average' galaxy. The errorbars plotted do only take into account the 
  error on 
  the measured velocity dispersion. Errors due to uncertainties in the 
  mass-velocity dispersion calibration are not taken into account 
  (see text for further details). The best fit parameters for the 
  mass-luminosity relation given by Eq.~\ref{eq:powerlaw} are given 
  for spirals and ellipticals, and plotted as solid lines.
  }
\label{fig:amag_mass}
\end{figure}

\subsection{Mass estimates for 2dFGRS primaries}

We now use the relation between the median velocity dispersion
of all central galaxies and halo mass,
shown in Fig.~\ref{fig:veldisp_mass}, to
directly convert the  measured satellite velocity dispersions into
estimates of halo masses.
In Fig.~\ref{fig:amag_mass}, we plot the estimated masses for 
primaries split by morphological type. As is to be expected,
we find in both cases that the inferred halo mass increases steadily 
with the absolute magnitude of the primary. 
With this mass calibration, 
we find that elliptical galaxies live in halos which are typically 3 to 
10 times more massive than spirals of similar b$_{\rm J}$ brightness. 

The scatter seen in Fig.~\ref{fig:veldisp_mass} indicates that 
the uncertainty in the mass calibration is large.
For a given halo mass, the scatter in velocity 
dispersion is around $\sim$~25 to $\sim$~40\%. Hence, considerable
care should be taken when using this relation to infer halo mass
from the measured velocity dispersion. 
In Fig.~\ref{fig:amag_mass}, we are
assuming that the satellite velocity dispersion measurement inferred from 
the 2dFGRS is in good agreement with the median of the `true' satellite 
velocity dispersion. In Fig.~\ref{fig:sig.mag}, we showed that for the
mocks the corresponding agreement is  
good, but not perfect. For the 
mocks we can use the spread between 16$^{\rm th}$ and 
84$^{\rm th}$ percentiles of the distribution as a guide to
the uncertainty in this calibration and hence on the
the systematic uncertainty in the calibration procedure. For 
an individual object, a systematic shift of $\sim$~75\% in the estimated 
mass is entirely acceptable, as this shift corresponds to the 1-$\sigma$ 
dispersion on the calibration relation. 

Using this calibration, we find the power-law fit relation 
\begin{eqnarray}
\rm \frac{M_{175}}{h^{-1}M_\odot} & = & \left(\frac{L_{\bjm}}{L_*}\right)^\alpha \left(\frac{M_{*}}{h^{-1}M_\odot}\right)
\label{eq:powerlaw}
\end{eqnarray}
between the  halo mass within 175~\hkpc, $M_{175}$,
and the $b_{\rm J}$-band luminosity of the primary, $L_{\bjm}$.
This can be written equivalently as
\begin{eqnarray}
\rm log_{10}\left[M_{175}/(h^{-1}M_\odot)\right] = \rm \alpha \
log_{10}\left[ L_{\bjm}/L_{*} \right] + \beta ,
\end{eqnarray}
where $L_*$ is the characteristic luminosity as given by the 2dFGRS 
Schechter luminosity function estimate of Norberg \etal~(2002),
 and $\rm \beta = log_{10}[ M_*/(h^{-1}M_\odot)]$, with $M_*$ the 
dynamical mass of an $L_*$ galaxy.
For elliptical galaxies, we observe a nearly linear relation 
between halo mass and luminosity, with $\alpha \simeq 1.0\pm0.4$ 
and $\beta \simeq 12.6\pm0.2$, i.e. a nearly luminosity independent 
mass-to-light ratio. This is in stark contrast 
with spiral galaxies, for which the relation between halo mass 
and primary luminosity is much steeper, with 
$\alpha \simeq 1.6\pm0.6$ and $\beta \simeq 11.8\pm0.3$. 
The errors on these best fit parameters are, as already mentioned, 
rather substantial. 


We note that the existence of a difference between the scaling
relations for ellipticals and spirals is independent of the 
calibration used. The calibrating relation between
halo mass and satellite velocity dispersion was assumed to
be independent of galaxy 
morphology and  hence the difference in the mass-luminosity
scaling relations is due entirely to the differences in the measured
luminosity - velocity dispersion relations.
However the magnitude of this difference does depend on 
the halo mass-velocity dispersion calibration used.

Finally, for a spiral galaxy like the Milky-Way
we estimate the halo mass within 175~\hkpc\ to be 
approximatively $\rm 3.5^{+4.0}_{-2.1} \times 10^{11}$~\hmsol,
whereas for an elliptical galaxy of similar b$_{\rm J}$ 
brightness we estimate 
its halo mass to be nearly eight times larger. We note the large 
statistical error associated to this mass estimate, which do not 
include any systematical error on the mass calibration relation used.

\subsection{Comparison with other mass estimates}

This statistical Milky-Way mass estimate needs to be compared with 
other Milky-Way mass estimate, using completely different 
techniques. When comparing with techniques independent 
of the value of the Hubble constant, we assume here 
$\rm H_0 = 70\,km\,s^{-1}\,Mpc^{-1}$, implying our estimate of the 
Milky-Way mass within 250 kpc becomes 
$\rm 5.0^{+5.7}_{-3.0} \times 10^{11}\,M_{\odot}$. Assuming a simple scaling 
relation for the mass enclosed at large radii (e.g. $\rm M(r) \sim r$, 
isothermal sphere), we estimate the mass within 100~kpc to be 
$\rm 2.0^{+2.3}_{-1.2} \times 10^{11}\,M_{\odot}$. 

From the dynamics of the Magellanic Clouds and the 
associated stream, Lin, Jones, \& Klemola (1995) estimate the mass of 
the Milky Way inside 100~kpc to (5.5$\pm$1)~10$^{11}$~\msol.
From the escape velocity and motions of satellite 
galaxies, Kochanek (1996) estimates the mass of the Galaxy inside 
100~kpc to be $(5-8) \times 10^{11}$~\msol. Using more recent kinematic 
information for Galactic satellites and halo objects, Sakamoto, 
Chiba, \& Beers (2003) derive an essentially model 
independent Galaxy mass estimate within $\sim$~50~kpc of 
5.5$^{+0.1}_{-0.2} \times 10^{11}$~\msol, which corresponds to 
$\sim$~10$^{12}$~\msol\ within 100~kpc. This mass, which is nearly 
twice as large as the one found by Lin \etal~(1995), seems to be 
confirmed by Bellazzini (2004), who uses the tidal radii of 
remote globular clusters (between 35 and $\sim$~200~kpc from 
the Galactic Centre) to provide constraints on the mass profile 
of the Milky Way, independently of kinematic data and yielding 
an enclosed mass of 1.3$^{+2.9}_{-1.0} \times 10^{12}$~\msol\ at 
$\sim$~90~kpc.

All these estimates are at least a factor of two larger than our
estimate and the more recent Milky-Way mass estimates are as much
as a factor 5 larger. There are several systematic effects that
could be contributing to this difference. Firstly, our estimate
is a statistical estimate of the mean mass for galaxies of a given
luminosity. The scatter about this mean relation is large
and ought to be taken into account. For instance,
the scatter of the halo mass satellite velocity dispersion 
calibration relation is 25\% for a given mass, which translated into 
the mass luminosity relation implies a typical 75\% scatter for a given 
luminosity. 
Secondly, our mass estimate for the Milky-Way is obtained 
using a power-law fit to the data, which therefore averages
data over a range of luminosities. If a power-law is not 
appropriate and only the data in a bin centred at the Milky-Way's
luminosity were used then the statistical error on our estimate
would be $\sim$~70\% larger.
Finally, according to the Milky-Way models of 
Kochanek (1996), the assumption of isothermal sphere between 
100~kpc to 250~kpc may not be  appropriate. Following 
his figure~7, the scaling is closer to $\rm M(r) \sim r^{\sim 0.6}$, 
implying a $\sim$~50\% increase in the estimated Milky-Way mass.

\begin{figure}
\centering
\epsfig{file=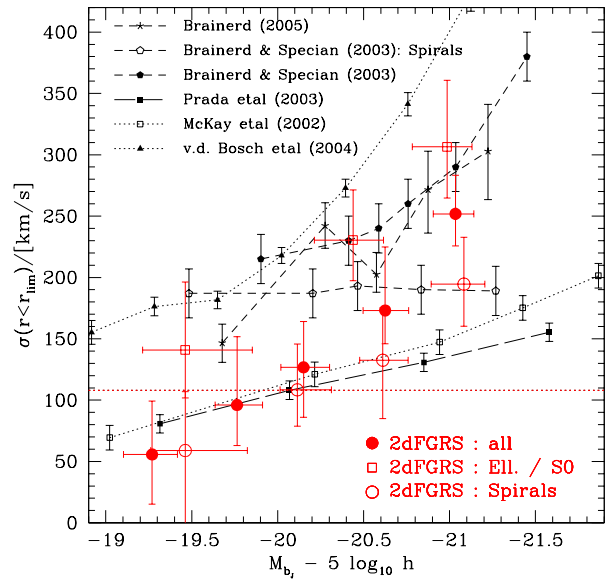,width=0.47\textwidth,clip=,}
\caption{Comparison of velocity dispersion estimates as function of
  absolute magnitude from recent analyses using similar techniques. For
  Prada \etal~(2003), McKay \etal~(2002a), Brainerd \& Specian (2003)
  and Brainerd (2005), the 
  quoted errorbars are upper limits on the errorbars shown in their
  respective analyses. For \vdbpapa, we show their final velocity dispersions 
  estimates, for their satellite sample optimised so as to have, according 
  to mock catalogues constructed using the conditional luminosity function, 
  the lowest amount of interlopers. Our measurements, within 175~\hkpc, 
  are the same as presented in Fig.~\ref{fig:sigma_amag_morph}. See 
  section~\ref{sec:discussion}, but also the 
  Appendix~\ref{sec:comp_sdss},~\ref{sec:brainerd} and~\ref{sec:vdbpap} 
  for further comments on this rather busy plot.}
\label{fig:sigma_comp}
\end{figure}

Taking these statistical and systematic issues into account, our 
mass estimate is compatible with those of Lin \etal~(1995) 
and Kochanek (1996), but still roughly a factor of 2-3 too small 
when compared to the recent estimates from Sakamoto \etal~(2003) and 
Bellazzini (2004). Clearly our statistical method is 
not optimal for inferring the mass of an individual 
object. A fairer comparison would be to compare our results
with other statistical mass estimators, like
McKay \etal~(2002) did when comparing with 
weak lensing estimates from SDSS (McKay \etal~2001).


\section{Comparison to similar methods}
\label{sec:discussion}

There have recently been several complementary studies that have
used similar techniques to stack satellites around isolated
primaries (McKay \etal~2002; 
Prada \etal~2003; Brainerd \& Specian 2003; van den Bosch \etal 2005;
Brainerd 2005). The results of these studies,
appropriately converted into \mbjlogh\ where necessary, are compared to
ours in Fig.~\ref{fig:sigma_comp}.
At first sight, one sees very large variation between the findings of 
the different authors. However one has to be very careful because
both the selection criteria and method of estimating velocities vary
considerably. A more detailed comparison between
the different analysis methods is presented in the Appendices. The
tests presented there show why some estimates differ, but also raise
new concerns over other estimates which on face value appear to agree
when in reality they probably should not. 

From the analysis presented in the Appendix~\ref{sec:comp_sdss}, 
we learn first that the agreement in the results from 
McKay \etal~(2002) and Prada \etal~(2003) is, as already claimed 
by Prada \etal~(2003), spurious. The selection criteria are
different enough so that when applied to mocks the results are 
expected to be significantly
different. It is clear from the paper of Brainerd \& Specian (2003)
that they have not subtracted the pairwise velocity measurement
uncertainty from their velocity dispersion estimate, while Brainerd
(2005) does take it into account. Despite this 
(see Appendix~\ref{sec:brainerd}), the results from
Brainerd \& Specian (2003) and Brainerd (2005) still do not agree with our
findings also extracted from the 2dFGRS. We believe this is due to
some source of extra contamination of their satellite sample,
but we are unable to reproduce their results.
Moreover, we find that the difference  between our measurements and
those of Prada \etal~(2003) is mostly due to the different selection
criteria used to define the samples.
The larger errors on our 2dFGRS velocity dispersion estimates
cannot be explained solely by the larger redshift measurement errors of
the 2dFGRS data. Using the data of Prada \etal~(2003), we find their
quoted errors to be approximatively 30-40\% smaller than those we
estimate.

As a final conclusion from this comparative work, we have to point out
that the large variety of results present in the literature reflects
mostly the large variation in the proposed methods of  both identifying
the isolated systems and also measuring the satellite velocity 
dispersions. All the proposed methods introduce biases and unless
one applies the same selection criterion to a set of realistic mock galaxy
catalogues one cannot quantify these systematic effects and the
interpretation of the results remain questionable.

\section{Conclusions}
\label{sec:conclusion}

We have developed, tested and applied a method to probe the properties
of extended dark matter haloes around bright galaxies. We do this by
carefully selecting isolated galaxies in the 2dFGRS and using their
faint satellites as tracers of the gravitational potential. By
stacking systems of similar primary luminosity to improve the
signal-to-noise, we estimate the satellite velocity
dispersion. Realistic mock galaxy catalogues, created from
cosmological N-body simulations populated with a semi-analytic galaxy
formation scheme, enable us to relate the measured velocity dispersion
of the satellite system to the velocity dispersion and mass of the
underlying dark halo.

Fig.~\ref{fig:sigma_amag_morph} shows evidence for the existence of
dark matter halos around typical galaxies. Our sample of satellites
probes the potential well of the primaries out to several hundred
kiloparsecs and demonstrates that the dark halos extend many times
beyond the optical radius of the primary. This is in agreement with
the current theoretical picture of galaxy formation in a cold dark
matter universe (e.g. White \& Frenk 1991, Kauffmann \etal~1993, Cole
\etal~2000). The satellite velocity dispersion increases with the
luminosity of the primary and is much larger for elliptical galaxies
than for spiral galaxies of similar b$_{\rm J}$ luminosity.  

The total extent of the dark halo is not constrained by our data. 
Although, the satellite distribution extends to $r_p
\sim$~375~\hkpc, most of the signal comes from within $r_p
\sim$~175~\hkpc. Within the errors, the velocity dispersion appears to
be constant within $r_p \sim$~175~\hkpc\ and $r_p \sim$~375~\hkpc. In
this range, the satellite velocity dispersion does not depend strongly
on the luminosity of the primary.  

Our mock catalogues allow us to calibrate the velocity dispersion-mass
relation for galaxies selected according to the isolation criterion of
our 2dFGRS sample. Fig.~\ref{fig:amag_mass} then indicates that
elliptical galaxies reside in haloes which are at least 4 times more
massive than spiral galaxies of similar b$_{\rm J}$ brightness. Galaxy
like the Milky-Way typical reside in dark matter halos of mass 
$\rm \sim 3.5^{+4.0}_{-2.1}\,10^{11}$~\hmsol\ within 175~\hkpc.

A key assumption in our analysis is that isolated galaxies of similar
luminosity reside in halos of similar mass. It is this that justifies
the stacking procedure. Our semi-analytic models allow us to test the
validity of this assumption. We find that in one of two semi-analytic
models that we have investigated, the \munichsam\ model, there is very
little scatter in the relation between central galaxy luminosity and
halo mass. A different semi-analytic model, however, that by Cole
\etal~(2000), predicts considerable scatter in this relation and this
introduces large errors in the halo properties inferred from a
stacking analysis. 
Mocks constructed from this model return an
increasing satellite velocity dispersion as a function of primary
luminosity which, however, deviates systematically from the velocity
dispersion of the host dark halos. 
The reasons behind this difference
in the galaxy formation models are not investigated in detail
here but it serves to illustrate that significant theoretical
uncertainties remain in the kind of analysis that we have presented
here.

\section*{Acknowledgements}

We thank Tim McKay for a very helpful referee's report. The 2dFGRS was
carried out using the 2 degree field facility on the 3.9m
Anglo\-Australian Telescope (AAT). We thank all those involved in the
smooth running and continued success of the 2dF and the AAT. During
the far too long course of this work, PN acknowledges funding from the Swiss
National Science Foundation (TMR PhD Grant), an ETH Zwicky Fellowship,
a PPARC fellowship at the IfA, as well as a visitor status at the ICC
and many stimulating discussions with Felix Stoehr, Simon White, Liang
Gao, Cristiano Porciani, Frank van den Bosch, Diego Lambas and
Francisco Prada. CSF acknowledges a Royal Society Wolfson Research
Merit award.

\appendix

\renewcommand{\theequation}{A\arabic{equation}}
\setcounter{equation}{0}  

\renewcommand{\thefigure}{A\arabic{figure}}
\setcounter{figure}{0}  

\section{Comparison with satellite velocity dispersion 
measurements from SDSS data}
\label{sec:comp_sdss}

The studies of McKay \etal~(2002) and Prada \etal~(2003) are based 
on different releases of SDSS data. In Fig.~\ref{fig:sigma_comp},
their results are shown 
by dot-connected open squares and by long-dashed connected filled
squares respectively. 
At first sight their results appear in rather good agreement with each other, 
and rather different to ours, which are shown by large filled circles 
(all primaries), large open circles (spiral primaries) and large open 
squares (elliptical/S0 primaries). However, it is essential to consider
the differences expected due to the 
differing satellite selection criteria. 

\subsection{McKay et al.: $\delta m_s = 1.5$ \& $\delta r_b = 2$~\mpc}
\label{sec:comp_mckay}

McKay \etal~(2002) (and also Prada \etal~(2003) with their 
sample 3\footnote{This sample is presented in their paper, but
is not the sample for which they derive their main results.})
use a much less stringent isolation criterion on the neighbourhood
of the primary. Our requirement for the primary to be at least 
8 times more luminous than any of its satellites is relaxed to just 
4 times brighter.  We recall that the role of this constraint is
to avoid including satellite systems in which the potential well
and hence the dynamics of the satellites is not dominated by the
the primary. To avoid this being the case it is necessary to ensure
that all the satellites are much less massive than the primary.
The recent mass-to-light ratio measurements of  Eke \etal~(2004)
indicate that for around 2~\lstar, which is the luminosity of 
our brighter primaries, a factor of 8 in luminosity corresponds to 
factor 10 in mass. However for lower luminosity primaries the
corresponding mass factor is smaller and this argues for having a 
large luminosity difference between satellite and primary.

Another difference between our standard isolation criterion and the one 
used by McKay \etal is the much larger outer exclusion radius. 
Our requirement is to have all galaxies 
within 1~\mpc\ be at least 0.8 magnitudes fainter than the primary, 
whereas McKay \etal require for the same magnitude difference a 
distance of 2~\mpc, which is certainly a more restrictive 
criterion.

\begin{figure}
\centering
\epsfig{file=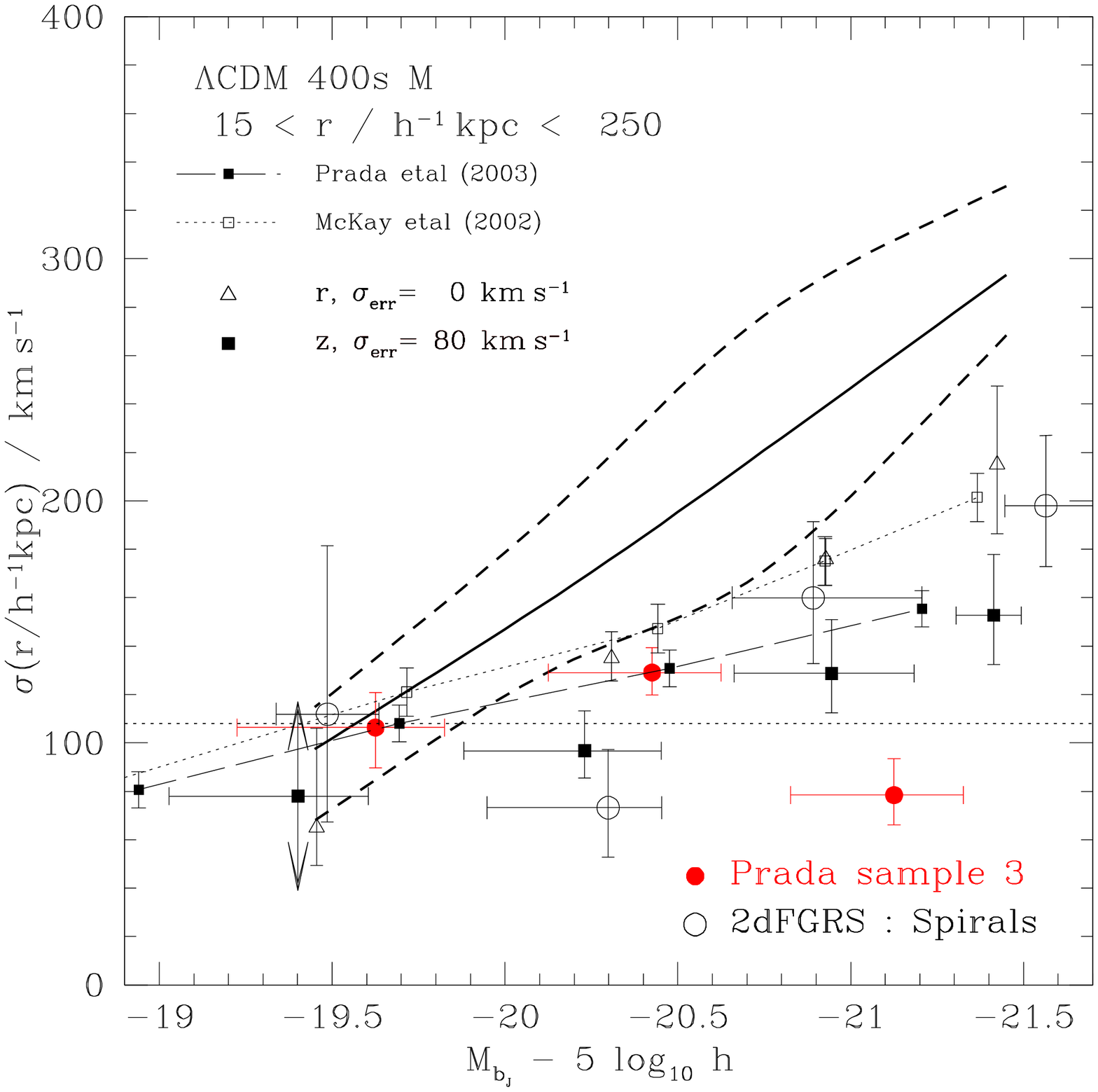,width=0.47\textwidth,clip=,}
\caption{Comparison of velocity dispersion estimates as function of
  absolute magnitude, adopting the same criterion as McKay
  \etal~(2002) and Prada \etal~(2003) in their sample 3. The labelling
  is the same as in Fig.~\ref{fig:sigma_comp}, with the exception that 
  we have added our mock catalogue results (open triangles and filled 
  squares for real and redshift space respectively) for this new 
  isolation criterion, where the velocity dispersion is measured for 
  satellites between 15 and 250~\hkpc\ using a cylinder depth of
  500~\kms. The filled large points correspond to our measurement done 
  on sample 3 of Prada \etal~(2003). Note that the connected filled squares 
  from Prada \etal~(2003) are estimates using the selection criterion 
  discussed in Fig.~\ref{fig:sigma_comp_prada}.}
\label{fig:sigma_comp_mckay}
\end{figure}

Therefore in order to make the appropriate comparison, we need to apply 
the same selection criterion as McKay \etal~(2002). 
Fig.~\ref{fig:sigma_comp_mckay} compares the results of McKay
\etal~(2002) with new estimates we have made from the 2dFGRS data
after adopting their selection criterion. Also shown in 
Fig.~\ref{fig:sigma_comp_mckay} are velocity dispersion estimates
we have made using sample 3 of Prada \etal~(2003).
Using the same selection criterion, we now find results which are in 
better agreement with the SDSS estimates, but with larger errorbars.
There are two reasons why are our errors are larger.
Firstly, the much larger velocity uncertainty of 2dFGRS 
redshift measurements (typically  85~\kms\ 
on an individual galaxy for the 2dFGRS, compared to 
less than 30~\kms\ for SDSS) causes the 
errors from the maximum 
likelihood method to be increased by 50~\%, for systems with 
a `true' satellite velocity dispersion of 100~\kms, 
instead of just 7~\% in the case of SDSS.
 Secondly, the errors quoted by Prada \etal~(2002) are 
intrinsically smaller than those we obtain when performing our maximum 
likelihood estimation on their sample 2. This result can be 
partially explained if Prada \etal have assumed Poisson statistics 
to derive their quoted errorbars.

Comparing the results of applying this relaxed selection criterion to
the mock catalogues, with the distribution of the underlying satellite
velocity dispersions shown by the heavy solid and dashed lines in
Fig.~\ref{fig:sigma_comp_mckay}, we see that there is a large bias. The
velocity dispersion recovered from the mocks lie outside the 16$^{\rm
th}$ to 84$^{\rm th}$ percentile band of the `true' underlying galaxy
velocity distribution. 
In our analysis of the samples with relaxed isolation criterion we
also note that for both the 2dFGRS data and the mocks 
the stacked satellite velocity distributions 
are no longer well fit by the `Gaussian plus a constant' model.
The samples look much more like 
the real space velocity distributions shown in Fig.~\ref{fig:fit_3magbin}, 
i.e. there is just an upper limit for the constant. In other words, 
with these satellite samples, fitting the 2dFGRS data with a 
`Gaussian plus a constant' is not appropriate. On 
the other hand, the satellite velocity distribution of sample 3 of 
Prada \etal is well fit by a `Gaussian plus constant'. This 
indicates some clear difference between the two satellite samples.

\begin{table*}
\caption{Properties of the combined NGP \& SGP satellite samples 
around bright galaxies, for two values of $N_{\rm viol}$ and for 
different selection criteria, as indicated by $\delta m_s$ and 
$\Delta V_s$. The McKay \etal selection criterion
correspond to $\delta m_s=1.5$, Prada 
\etal to $\delta m_s=2.0$ and ours to $\delta m_s=2.2$. All satellites 
are within 250~\hkpc\ from the primary. The columns 
labelled `large systems' indicate how many primaries and satellites 
are found in systems with 6 or more satellites within 250~\hkpc. The 
last 3 columns show the total number of satellites within two cylindrical 
shells, where $r_{\rm p}$ is expressed in \hkpc. These numbers include 
the satellites belonging to `large systems'.}
\begin{tabular}{cccccccccc}   
\hline
\multicolumn{3}{c}{Selection Criteria} & N$_{\rm prim}$ & N$_{\rm sat}$ & \multicolumn{2}{c}{large systems} & \multicolumn{3}{c}{N$_{\rm sat}$} \\
$\delta m_s$ & $\Delta V_s$ & N$_{\rm \rm viol}$ & & & N$_{\rm prim}$ & N$_{\rm sat}$ & $15<r_{\rm p}<90$ & $90<r_{\rm p}<250$ & $15<r_{\rm p}<250$ \\
\hline
2.2 & 600 & 0 & 299 & 425 & 3 & 21 & 133 & 289 & 422 \\
2.0 & 600 & 0 & 513 & 754 & 7 & 52 & 244 & 499 & 743 \\
1.5 & 600 & 0 & 233 & 370 & 3 & 18 & 102 & 267 & 369 \\
\hline
2.2 & 1200 & 0 & 270 & 402 & 4 & 29 & 125 & 275 & 400 \\
2.0 & 1200 & 0 & 482 & 747 & 9 & 68 & 232 & 505 & 737 \\
1.5 & 1200 & 0 & 200 & 339 & 6 &  41 & 92 & 246 & 338 \\
\hline
2.2 & 600 & 4 & 490 & 684 & 4 & 30 & 217 & 450 & 677 \\
2.0 & 600 & 4 & 820 & 1160 & 8 & 58 & 359 & 786 & 1145 \\
1.5 & 600 & 4 & 475 & 741 & 7 & 48 & 200 & 540 & 740 \\
\hline
2.2 & 1200 & 4 & 415 & 610 & 6 & 45 & 186 & 420 & 606 \\
2.0 & 1200 & 4 & 751 & 1121 & 12 & 87 & 322 & 785 & 1107 \\
1.5 & 1200 & 4 & 365 & 627 & 11 & 78 & 163 & 463 & 626 \\
\end{tabular}
\label{tab:stats_2}
\end{table*}

\subsection{Prada et al.: $\delta m_b = \delta m_s = 2.0$ 
\& $\delta r_b = 0.5$~\mpc}
\label{sec:comp_prada}

\begin{figure}
\centering
\epsfig{file=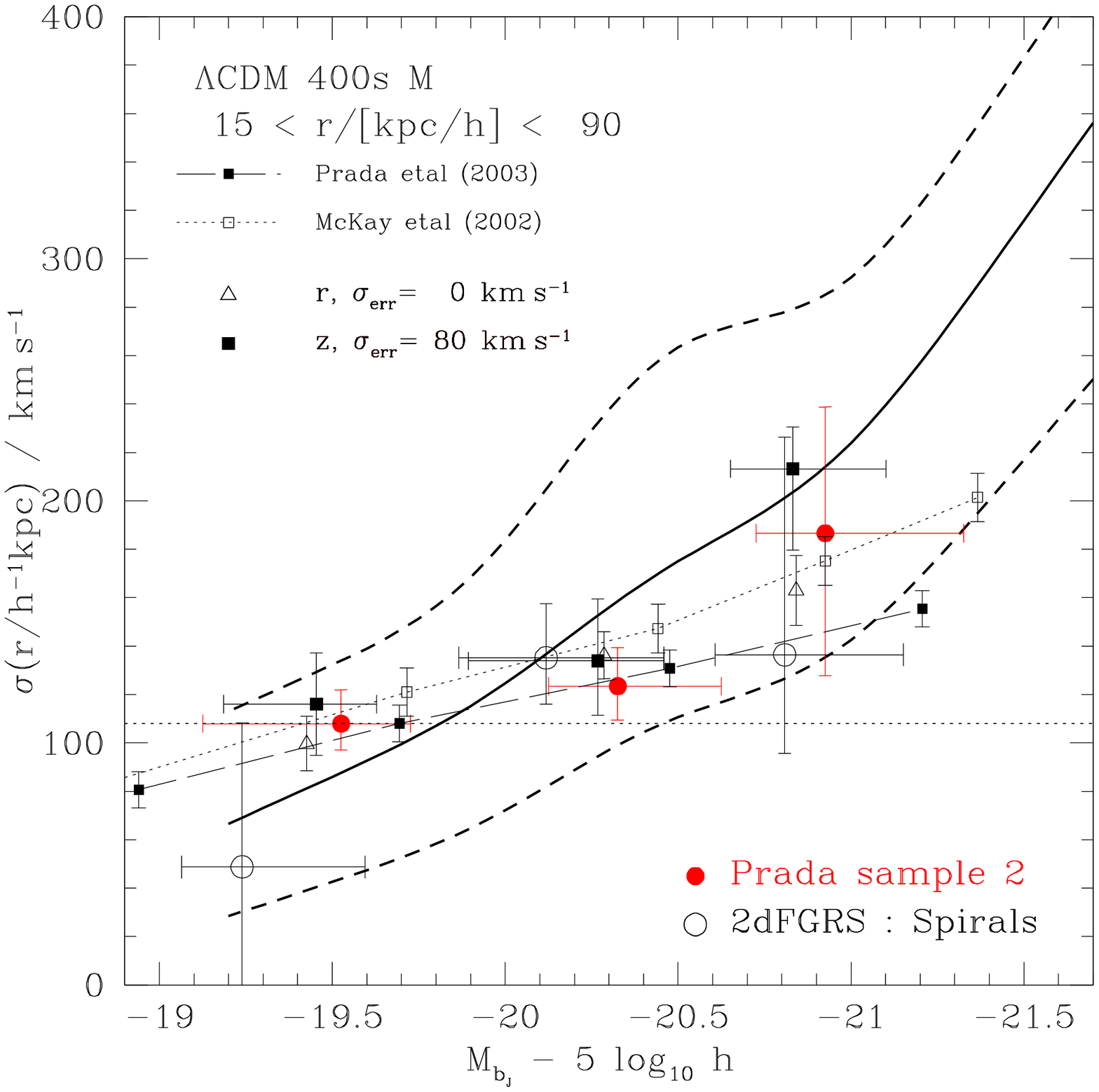,width=0.47\textwidth,clip=,}
\caption{Comparison of velocity dispersion estimates as function of
  absolute magnitude, adopting the same criterion as used by Prada
  \etal~(2003). The labelling is the same as in
  Fig.~\ref{fig:sigma_comp}, 
  with the exception that we have added our mock catalogue results 
  (open triangles and filled squares for real and redshift space 
  respectively) for this new isolation criterion, where the velocity 
  dispersion is measured for satellite between 15 and 90~\hkpc\ using a 
  cylinder depth is 500~\kms. The filled large points correspond to our 
  measurement done on sample 2 of Prada \etal~(2003). Note that the 
  dot-connected open squares from McKay \etal~(2002) are obtained with 
  the selection criterion discussed in Fig.~\ref{fig:sigma_comp_mckay}.}
\label{fig:sigma_comp_prada}
\end{figure}

The selection criterion proposed by Prada \etal~(2003) are 
more conservative than those of McKay \etal~(2002), as they require a factor 
of $\sim$~6 in luminosity between the primary and any of the 
surrounding galaxies. However, for the spatial isolation of the 
primary, Prada \etal~are much less restrictive with an 
outer isolation radius, $\delta r_b$, of 0.5~\mpc, compared to 2~\mpc\ 
for McKay \etal~(2002). Compared to our choice, 
this criterion is less restrictive, 
especially as we require within $\delta r_s = 0.5$~\mpc\ a factor 
of 8 in luminosity difference, and within $\delta r_b = 1$~\mpc\ 
the primary to be at least twice as luminous as any other galaxy. 
However, this 
is not the only difference between the two SDSS satellite analyses.
The biggest difference is that Prada \etal fit for the 
satellite velocity dispersion and interloper fraction separately
in each bin of projected radius. Thus they allow both the
interloper fraction and velocity dispersion to vary with projected
radius.

Including this extra freedom in their velocity dispersion 
calculations gives rise, as we show in Fig.~\ref{fig:sigma_comp_prada}, 
to a substantial difference in the velocity dispersion estimates, 
compared to what is found in Fig.~\ref{fig:sigma_comp_mckay}, 
using the analysis method proposed by McKay \etal~(2002). From 
Fig.~\ref{fig:sigma_comp_prada}, we conclude that satellite 
velocity dispersion estimates from 2dFGRS and SDSS (using 
sample 2 of Prada \etal~2003) agree with each other and that the 
criterion and method used seem to be able to recover (within 
its large uncertainties) the underlying velocity dispersion, 
which is not the case for the method of McKay \etal~(2002). 
Finally we are able, but with larger errors, to recover
the velocity dispersion estimates found by Prada \etal~(2003). 
Only for the brightest galaxies is there any indication of a
discrepancy, and this is mainly related to error estimates from
Prada \etal~(2003) which for those samples are clearly much smaller 
than we find.

Regarding the statistics of the isolated satellite systems, we note
that adopting the selection criterion of Prada \etal~(2003) causes the
number of systems with at least 6 satellites within 250~\hkpc\ to
increase drastically. This statistic is similar to the one we used
before, i.e. at least 9 satellites within 400~\hkpc, to identify and
remove small groups from our sample.  Table~\ref{tab:stats_2} compares
these statistics for the 2dFGRS data for our original selection
criterion, those of Prada \etal~(2003) and those of McKay \etal\
(2002). Typically the Prada \etal~(2003) selection criterion results
in twice as many 'large systems' as for our standard selection
criterion, independent of the value of N$_{\rm viol}$ and $\Delta V_s$
used.  Hence the sample of primaries might not be as well isolated as
one hopes. Interestingly, as long as N$_{\rm \rm viol}=0$, the
selection criterion of McKay \etal\ are very comparable to ours as
regards the number of `large systems' identified.  Nevertheless, we
note that in all cases the fraction of satellite galaxies that are in
these `large systems' is always small.

\subsection{Results from mocks}

With the comparisons of the two previous subsections, we have not 
been able to really quantify the difference between the proposed 
isolation criteria, even though our results hint towards the fact 
that the isolation criterion proposed by McKay \etal is probably 
the least appropriate of the three considered and that the one
by Prada \etal seems to identify slightly more `large systems' 
than ours. Therefore, in this section, we use the mocks 
and address the issue of the radial dependence of the satellite 
velocity dispersion, which Prada \etal found clear evidence 
for in their data.

\begin{figure*}
\centering
\plotthree{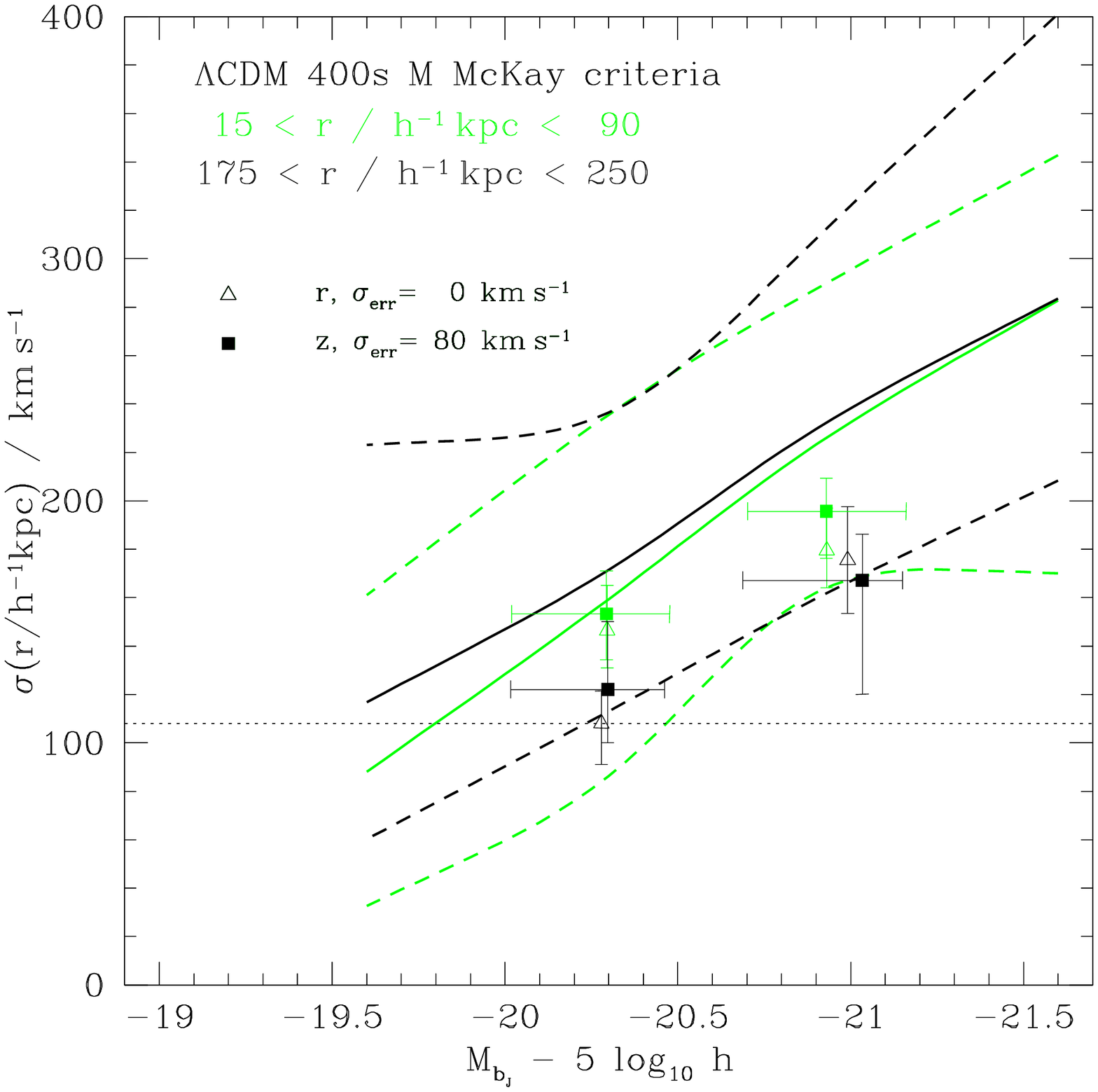}{0.356}{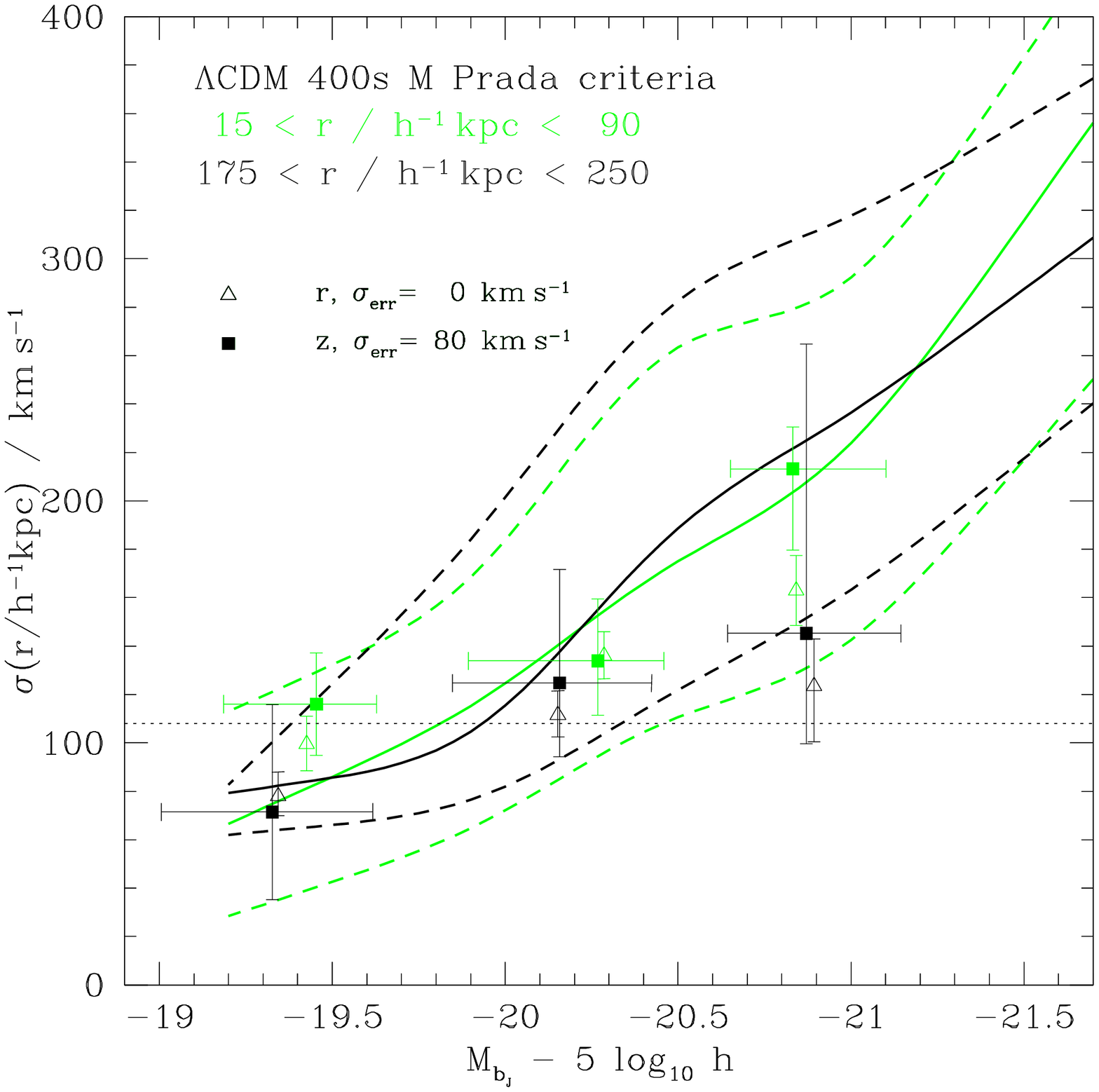}{0.316}{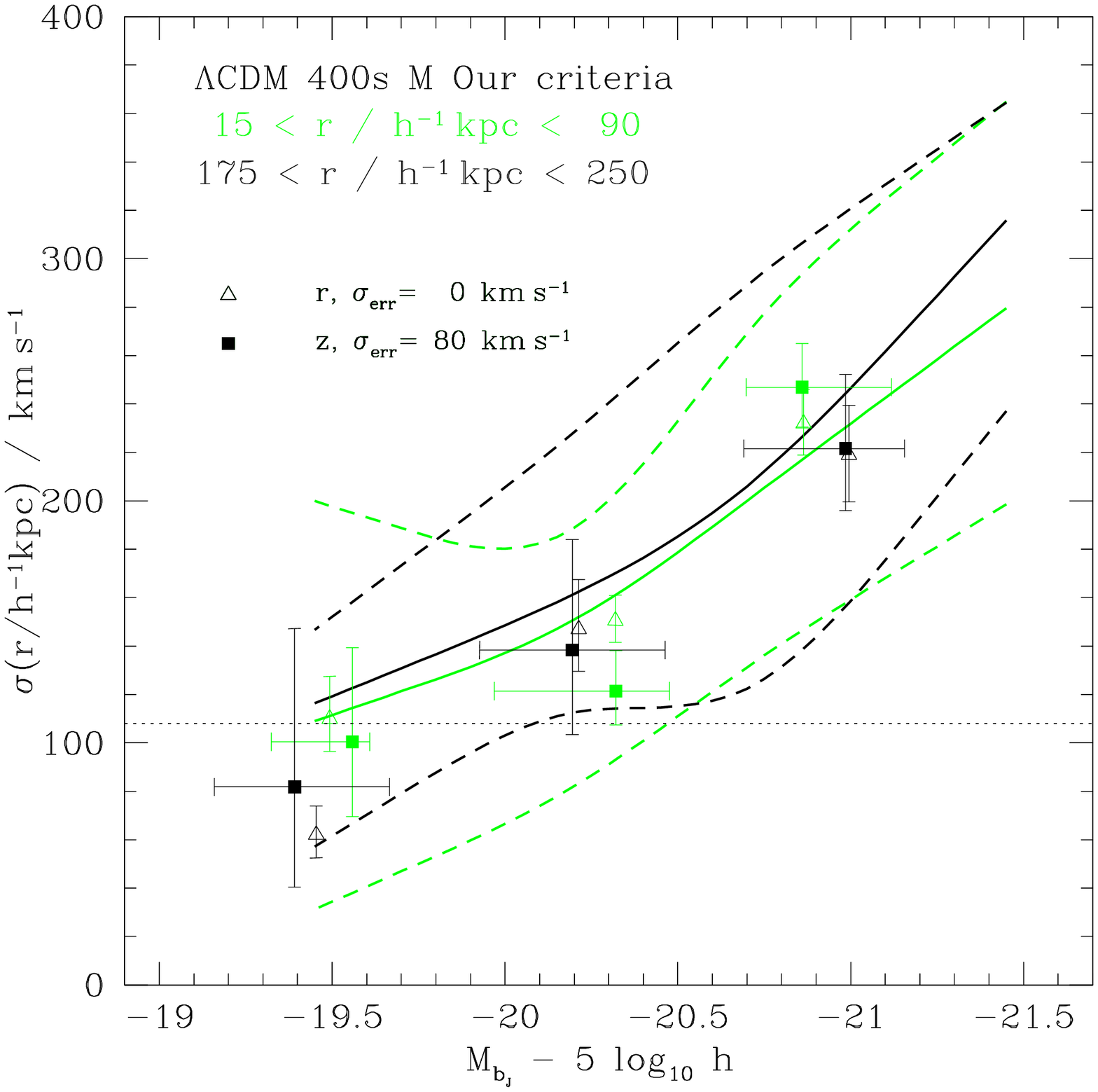}{0.316}
\caption{Mock results using the isolation criteria of McKay \etal (left 
 panel), of Prada \etal (centre), and of this paper (right panel). The 
 symbols with errorbars correspond to the real (open triangle) and 
 redshift (filled square) space satellite velocity dispersion measurements 
 around mock primaries identified using different selection criteria. The 
 lines correspond to the median (solid) and 16$^{\rm th}$ and 84$^{\rm th}$ 
 percentiles (dashed) of the underlying primary satellite velocity 
 dispersion. The different shadings correspond to two different 
 cylindrical shells, whose ranges are given in the panel.}
\label{fig:sigma_comp_4}
\end{figure*}

In Fig.~\ref{fig:sigma_comp_4} we show a comparison between the 
different selection criteria used by McKay \etal (left panel), 
Prada \etal (central panel) and us (right panel). The aim is to 
show how well each of these different selection criteria 
succeed in recovering the 'underlying' satellite velocity 
dispersion, for which the median is given by the solid lines, 
and the dotted lines represent the 16$^{\rm th}$ and 84$^{\rm th}$ 
percentiles of the satellite velocity distribution as measured from 
the simulation cube. For comparison purposes, we have chosen to present 
in each panel the results within two different projected radii, as 
in Prada \etal~(2003): $15<r_{\rm p}/\hkpc<90$ and 
$175<r_{\rm p}/\hkpc<250$.

First of all we note that the different criteria are not all as 
successful in recovering the underlying satellite velocity dispersion. 
Clearly the one proposed by McKay \etal (i.e. left panel of 
Fig.~\ref{fig:sigma_comp_4}) is the least successful, as it 
systematically gives an underestimate of the satellite 
velocity dispersion. This is especially true for the outer radial bin, 
for which the measured satellite velocity dispersion is barely within 
the 16$^{\rm th}$ percentile of the underlying satellite velocity dispersion 
distribution. Moreover we note that for the mocks there are not enough 
faint primary systems for which this satellite criterion is satisfied, 
explaining why no velocity dispersion measurement is given for primaries 
fainter or equal to \mstar. 
Interestingly, this is not true for the real data, for which 
there are still several faint primary satellite systems satisfying 
the isolation criterion. With larger number statistics, this could 
be a potential way of constraining certain galaxy formation models, 
an approach followed up by \vdbpapb.

The middle and right panels of Fig.~\ref{fig:sigma_comp_4} look more 
similar, in the sense that they both recover, within the errors, the 
`underlying' satellite velocity dispersion, for both projected radial 
bins. However, three small differences can be noted. Firstly, the 
velocity dispersion errors around bright primaries are much larger using the 
isolation criterion of Prada \etal~(2003). We note that the consistency
between  
expectation and measurement is only reached due to large 
non-symmetrical errors. Secondly, the median satellite velocity 
dispersion in the outer projected radial bin is less accurately 
recovered with the isolation criterion of Prada \etal~(2003). Indeed,
both real 
and redshift space measurements have a tendency to predict a much 
flater luminosity-velocity dispersion relation than the underlying one. 
Thirdly, the real space measurements obtained with the isolation 
criterion of Prada \etal~(2003) severely underestimate the true 
underlying velocity dispersion, especially for the brighter primaries.

Finally we note that, for our mock catalogue, there is virtually no 
radial dependence of the satellite velocity dispersion on the choice
of radial shell (the distributions indicated by the
bold and shaded lines are all in very good agreement 
with each other).
It is worth pointing out that of the three proposed isolation criteria,
only the one proposed by Prada \etal hints at a radial dependence 
of the satellite velocity dispersion. The trend is definitively 
weak and would be insignificant if
if the errors happen to be underestimated, or even just assumed to 
be symmetrical.
The fact that this can happen is a potential worry for the claim made
by Prada \etal~(2003) for a radial dependence of the 
satellite velocity dispersion.

\subsection{Conclusions for SDSS comparison}

We now have to address the question of whether our claim, that the
isolation criteria of Prada \etal and McKay \etal are more relaxed than
our standard ones, is consistent with the differences found between 
the estimated velocity dispersions. 
Looking at the systematic difference between 
velocity dispersion measurements made with the two criteria, what we 
find is quite counter intuitive. One would probably expect, 
that relaxing the isolation criterion would result in measuring 
larger satellite velocity dispersions, whereas the opposite is found.
This can probably related to a third difference between 
our nominal satellite selection criterion and the one used by Prada \etal: 
the length of the cylinder, within which satellite galaxies need to 
reside in order to be considered in the velocity dispersion estimate. 
Using Monte-Carlo realizations of satellite samples\footnote{i.e. with 
same number of systems as in the 2dFGRS, with similar distribution 
of satellites per system, and with the inclusion of an intrinsic 
maximal 10 to 20~\% variation in the background and in the underlying 
velocity dispersion, so as to mimic to some extent that all systems are 
not exactly identical.} drawn from a `Gaussian plus a constant' 
velocity distribution, we see that if one uses exactly the same 
velocity criterion as in Prada \etal (i.e. $\Delta V_s = 500$~\kms), 
one starts to systematically underestimate the velocity dispersion of 
systems with intrinsic velocity dispersion larger than 180~\kms. At 
the same time, one systematically overestimate the background for 
those systems. Taking into account the velocity errors, this 
translates, in the case of the 2dFGRS, to systems which are best fit 
by a velocity dispersion of around or larger than $\sim$~210~\kms, 
corresponding to all primaries slightly brighter than \mstar\ 
(see e.g. Fig.~\ref{fig:fit_3magbin}). On those grounds, we motivate 
therefore the use of a deeper cylinder than Prada \etal in order 
to accurately measure the velocity dispersion of slightly larger 
systems. Our Monte-Carlo approach shows that we could have adopted a 
limiting velocity difference of 900~\kms\ for spiral primaries, 
whereas there is a need, for elliptical primaries, to go out 
to $\sim$~1200~\kms, in order to appropriately sample the satellite 
velocity distribution. 

Therefore, most of the discrepancy between our analysis and the two using 
SDSS data reside in the different selection criteria used. From our 
results using the mock catalogues, from our Monte-Carlo simulations 
and from statistics of large systems discovered by the 
isolation criteria, there is a hint in the direction that the relaxed 
isolation criteria used by McKay \etal and Prada \etal are not as 
appropriate for finding dynamically isolated systems as our more 
stringent isolation criterion. Nevertheless we have to point out that 
the data is not yet good enough to be able to fully discriminate between 
the methods chosen. 

Finally, a closer inspection of the two SDSS works shows that their 
findings are slightly different, something which was 
already pointed out in the analysis of Prada \etal~(2003). Indeed the 
outer radii within which the velocity dispersions are measured are 
very different. As Prada \etal claim a strong dependence of the 
satellite velocity dispersion on radius, the agreement seen in 
Fig.~\ref{fig:sigma_comp} is not as good as it looks. On this last point, 
we would like to add, that we are not able, with our isolation 
criterion applied to the 2dFGRS, to detect such a signal. We know that the 
behaviour for ellipticals and spirals is rather different as function 
of luminosity, and therefore it could be legitimate to ask whether 
the effect seen by Prada \etal as function of projected radius could be 
due to a change in their sample mix as function of luminosity. Indeed 
with our findings, for galaxies of similar brightness, ellipticals 
will reside in much larger haloes than spirals. Hence stacking galaxies 
together irrespective of their morphological type, as done by 
Prada \etal~(2003), could give rise to a velocity dispersion which 
depends on the radius within which it is measured. With our samples we are 
not able to reliably examine this issue, as it requires the samples to 
be split by morphological type, luminosity and projected 
radius. The only conclusion we can draw from our samples is that we observe 
a trend indicating that satellite velocity dispersion measurements 
of galaxies residing in the range $175\,\le\,r_p\,/\,$\hkpc$\,\le\,375$ 
do not contain much information. This is in perfect agreement with the 
fact that the measured satellite velocity dispersions within 375~\hkpc\ 
are identical, to within the errors, to those measured 
within 175~\hkpc.

\section{Comparison with Brainerd \& Specian and Brainerd }
\label{sec:brainerd}

Regarding a comparison with the measurements of Brainerd \& 
Specian (2003), for which in Fig.~\ref{fig:sigma_comp} 
ellipticals and spirals are shown by dashed-connected 
filled and open pentagons respectively, we first need to 
point out that they have used a similar selection criterion 
to the one proposed in Prada \etal~(2003). For that 
reason, we expect, as explained above, to find differences 
between their results and our standard ones. However, like we 
did for Prada~et al., we should be able to recover their results by
assuming the same selection criterion. 

For their sample of satellites around spiral primaries, it is 
impossible to recover their results for the following reasons. 
Firstly, they have 
forgotten to subtract the {\it rms} velocity measurement errors in 
quadrature, which in their case are of the same order of 
magnitude as ours, i.e. $\sim$~110~\kms\ (as they use data from 
the 2dFGRS 100k release). Secondly, in some way, the 
isolation criterion they have applied have to be wrong, as it is 
impossible to understand how they initially find an 
isolated system with more than 605 satellites. This is even 
larger than the largest galaxy cluster found in the complete 
2dFGRS by Eke \etal~(2004). In all likelihood they must have 
forgotten to deal with effects due to 2dFGRS 100k window 
function, which is extremely patchy, and hence very unsuitable 
for this type of study. Similarly, they have probably not used 
the full photometric input catalogue to reject systems 
which were not fully observed. For these reasons, their 
velocity dispersion measurements around spiral galaxies, from
data which is a subsample of what we used in this analysis, 
are strongly erroneous. 

Regarding their subsamples of elliptical primaries, there is 
no reason not to believe that they are affected by the same 
problems as their sample of spiral galaxies. However, due to 
the fact that the intrinsic velocity dispersion of those systems 
is much larger, forgetting to subtract in quadrature the velocity 
errors does not influence the results by more than 10 to 15\% 
systematically. On the other hand, we believe, due to their 
problems with the isolation criterion, that the errors they quote 
on the satellite velocity dispersion around elliptical primaries 
is probably underestimated, and that the very strong trend 
with luminosity is too large. 

Finally, we note that we have not, at all, been able to 
reproduce their results using either the full 2dFGRS sample nor 
the 2dFGRS 100k release sample. We suspect therefore some of the 
above mentioned problems to be the cause of these difference.

Recently, Brainerd (2005) made a new satellite analysis using 
the full 2dFGRS survey. Like for the Brainerd \& Specian (2003) work,
we are unable to reproduce in detail their findings, especially 
for faint primaries.

\section{Comparison with van den Bosch et al.}
\label{sec:vdbpap}

Finally, the results from \vdbpapa, shown by dot-connected filled
triangles in Fig.~\ref{fig:sigma_comp}, clearly show that, with an
isolation criterion that is too relaxed, the proposed method no longer
finds a majority of dynamically isolated systems. Hence their criteria,
as already discussed in their paper, were not intended and should not
used for selecting systems for dynamical studies.

We note that the analytic method presented in \vdbpapa\ for the
luminosity - velocity dispersion relation, derived for their
conditional luminosity function model, is in very good agreement with
our measurements of that same relation from our semi-analytic
mocks. Moreover, applying our isolation criteria to \vdbpapa\'s
conditional luminosity function mocks, we recover the underlying 
luminosity - velocity dispersion relation to great accuracy.
This is a very strong consistency test for two completely 
different sets of models, constructed and constrained by different
mechanisms.

\end{document}